\documentclass[10pt,journal,compsoc]{IEEEtran}
\ifCLASSOPTIONcompsoc
  \usepackage[nocompress]{cite}
\else
  \usepackage{cite}
\fi

\usepackage{graphicx}                
\DeclareGraphicsExtensions{.pdf,.png,.jpg,.jpeg,.eps} 
\graphicspath{{figures/}{pictures/}{images/}{./}} 
\usepackage{microtype}                 
\PassOptionsToPackage{warn}{textcomp}  
\usepackage{textcomp}                  
\usepackage{mathptmx}                  
\usepackage{times}                     
\usepackage{cite}                      
\usepackage{tabu}                      
\usepackage{booktabs}                  
\usepackage{multirow}
\usepackage{url}
\usepackage{amssymb}
\usepackage{amsmath}
\usepackage{float}
\usepackage{wrapfig}
\usepackage{subfigure}
\usepackage{color}
\usepackage[ruled, vlined]{algorithm2e}

\usepackage{paralist}
\let\itemize\compactitem
\let\enditemize\endcompactitem
\let\enumerate\compactenum
\let\endenumerate\endcompactenum

\usepackage[table,xcdraw]{xcolor}
\usepackage{stfloats}

\ifCLASSINFOpdf
\else
\fi
\begin{document}
%
\title{Federated Visualization: A Privacy-preserving Strategy for Aggregated Visual Query}
%
%
%
%

\author{Wei~Chen,~Yating~Wei,~Zhiyong~Wang,~Shuyue~Zhou,~Bingru~Lin,~and~Zhiguang~Zhou
\IEEEcompsocitemizethanks{
\IEEEcompsocthanksitem W. Chen, Y. Wei, and Z. Wang are with The State Key Lab of CAD \& CG, Zhejiang University, Hangzhou, Zhejiang 310058, China.\protect\\
E-mail: \{chenvis, weiyating, zerowangzy\} @zju.edu.cn\protect\\
\IEEEcompsocthanksitem S. Zhou, and B. Lin are with the Alibaba Group, Hangzhou, Zhejiang 310058, China.\protect\\
E-mail: \{tangmao.zsy, bingru.lbr\} @alibaba-inc.com\protect\\
\IEEEcompsocthanksitem Z. Zhou is with Zhejiang University of Finance and Economics, Hangzhou, Zhejiang 310018, China.\protect\\
E-mail: \{zhgzhou1983@163.com\}@163.com\protect\\
\IEEEcompsocthanksitem Wei Chen is the corresponding author.
}
}

%
%

\markboth{Journal of \LaTeX\ Class Files,~Vol.~14, No.~8, August~2015}%
{Shell \MakeLowercase{\textit{et al.}}: Bare Demo of IEEEtran.cls for Computer Society Journals}
%



\IEEEtitleabstractindextext{%

\begin{abstract}
  We present a novel privacy preservation strategy for aggregated visual query of decentralized data. The key idea is to imitate the flowchart of the federated learning framework, and reformulate the visualization process within a federated infrastructure. The federation of visualization is fulfilled by leveraging a shared global module that composes the encrypted externalizations of transformed visual features of data pieces in local modules. We design two implementations of federated visualization: a \textit{prediction-based} scheme, and a \textit{query-based} scheme. We demonstrate the effectiveness of our approach with a set of visual forms, and verify its robustness with evaluations. We report the value of federated visualization in real scenarios with an expert review.
\end{abstract}

\begin{IEEEkeywords}
Privacy-preserving visualization, federated visualization, decentralized visualization
\end{IEEEkeywords}}

\maketitle

\IEEEdisplaynontitleabstractindextext

%
\IEEEpeerreviewmaketitle

\section{Introduction}

\maketitle

In big data era, a wide array of data are being generated all the time and scattered over different data holders. It is a significant task to closely connect data holders and conduct collaborative data analysis for deeper insights and comprehensive conclusions. For example, the combination disease materials and cases collected in different locations and a medical institution is quite important for the analysis of infectious diseases~\cite{plis2016coinstac,saha2017see}. Similarly, urban data analysis demands the integration of mobility information acquired from subway, taxis and automobiles~\cite{oksanen2015methods,wang2018user}.

In visualization community, two kinds of methods are commonly used for collaborative data analysis including data-intensive visualization and distributed visualization. Conventional client/server based visualization and web-based visualization take a data-intensive mode, in which datasets are assembled, processed and visualized in a main server~\cite{chen2017vaud,zhou2018visual}. Alternatively, distributed visualization~\cite{brodlie2004distributed} employs a decentralized mode, in which datasets and tasks are divided into pieces over various clients, and required information can be transmitted among clients.

In the both modes, raw datasets and their processed results are allowed to be transmitted among clients, and users can easily aggregate and analyze the datasets at different stages. However, there is a great risk of privacy leakage in the entire course of data processing. For example, in the data-intensive mode, users can directly get the dataset from a client on the server side. In the distributed mode, users can reversely infer the data information of a client through additional data information~\cite{dasgupta2014opportunities}. Thus, it is urgent to provide a feasible multi-source data visualization method without privacy leakage.

Privacy protection is a classic proposition in the field of visualization. One recent trend in building big data infrastructures is the privacy awareness, as witnessed by numerous literature on security and privacy~\cite{wang2017utility,wang2018graphprotector}. Though privacy-aware visualization has been explored in the visualization community, existing works emphasize on protecting data leakage within a centralized visualization process~\cite{dasgupta2011adaptive,wang2017utility}. There are still a few challenges for tackling privacy issue in the decentralized visualization: C1. A conventional visualization pipeline consists of multiple steps, including data processing, transformation, visual mapping, and user interaction. Data leakage can take place in each step in decentralized mode. Data encryption should be thoroughly incorporated as a necessary component. C2. For the reason of privacy protection, data in local modules should be kept locally, and only desensitized information can be transmitted to the global module. Thus, to generate faithful visualization based on composed information in global module is quite different from that in centralized mode. C3. It is important to evaluate the validity and usability of decentralized visualization in the protection of privacy and collective analysis based on multi-source datasets. Multiple perspectives like encryption, privacy, and effectiveness should be considered different from centralized visualization.

In the field of data mining, facing the challenges of data privacy in distributed machine learning, federated learning employs a shared global model to federate the learned local models that run only in individual clients, and uses a variety of encryption schemes in the data transmission process [10-12]. Inspired by the idea of federating learned features rather than raw data, we design a federated aggregation visualization strategy, which divides the tasks of data transformation and visual mapping into pieces in clients (local modules) and then composes the secured results in a server (a shared global module). We propose a framework for privacy-preserving decentralized visualization, in which federating learned features are integrated into the conventional visualization pipeline~\cite{wang2016survey}.
A data encryption scheme is employed to transmit data between the server and clients. To localize the underlying data and visualization tasks, we design a two-stage pipeline: prior to the composition of encrypted visual features or parameters in the server, each client locally performs data transformations and visual encodings with its associated data(C1). Specifically, we implement two federated aggregation visualization schemes for different data analysis scenarios: a \textit{query-based} scheme and a \textit{prediction-based} scheme. The \textit{query-based} scheme computes encrypted and specified range of visual features locally and composes them globally, through which users can quickly obtain target visualization in a short response time. The \textit{prediction-based} scheme builds a prediction model to encode visual features computed from local data, and decode them in the global module. It can realize the analysis of all data scenarios, of course requires more time to train the model to fit the data distribution of each client (C2). Moreover, we develop a federated aggregation visualization framework integrating different federated visualization schemes, a rich set of visual cues and interactions, enabling users to evaluate the effectiveness and practicality of our federated visualization strategy in the protection of privacy and collaborative analysis of multi-source data (C3). 

In summary, the main contributions of our work are:
\begin{itemize}
\item a novel federated visualization framework for visualizing aggregated visual query of decentralized data. To our best knowledge, our approach is the first attempt to tackle data privacy issues in a decentralized visualization framework. 
\item two implementation schemes for different data analysis scenarios.
\item evaluation and verification based on real world data.
\end{itemize}

The rest of this paper is organized as follows. Section 2 reviews the related work. Section 3 explains our new strategy. Implementations and visual exploration are elaborated in Section 4. Section 5 presents the evaluations. We discuss the limitations and future work in Section 6 and conclude this paper in Section 7.
\section{Related Work}

\subsection{Privacy-preserving Visualization}

Privacy is the right of individuals to take complete control of their information and decide when, how, and to what extent this information is shared with others~\cite{agrawal2002hippocratic}.
Visualization is regarded as an effective means to make the data or data features easily recognizable and interpretable. 
Therefore, privacy issues of visualization exist not only in data processing but also visualization itself. 
In the field of privacy-preserving data processing, two types of privacy models, syntactic anonymity~\cite{sweeney2002k, machanavajjhala2007diversity, li2007t, cao2012publishing} and differential privacy~\cite{soria2013differential}, are commonly used to address privacy issues. The former one is mainly used for privacy-preserving data publishing, the latter is designed to anonymize query responses. 
Pioneered works on privacy-preserving visualization~\cite{dasgupta2011adaptive} generally leverage established privacy preservation schemes like \textit{k}-anonymity~\cite{sweeney2002k}, \textit{l}-diversity~\cite{machanavajjhala2007diversity}, \textit{t}-closeness~\cite{li2007t} to protect information leakage in visualizations. For instance, syntactic anonymization is employed to protect the information exposed in parallel coordinates~\cite{dasgupta2011adaptive}. A privacy-preserving diversity method (ppDIV)~\cite{oksanen2015methods} is presented to avoid disclosure of location privacy from trajectory heatmap. Distinctive visual interfaces are designed to depict and reduce the leakage risk in visualizing event sequence datasets~\cite{chou2016privacy,chou2019privacy}, and tabular data~\cite{wang2017utility}. In general, there is a trade-off between the privacy gain and loss of utility. It has demonstrated to be favorable to support visual understanding of the quantitative relationship between privacy parameters and vulnerable visualization configurations~\cite{dasguptaguess}.

Note that, above-mentioned works do not consider the disclosure of privacy in data transmission process, and are naturally not suitable for decentralized visualization.  

\subsection{Distributed Visualization}

Distributed visualization ~\cite{brodlie2004distributed,chen2020federated} increases the scalability of visualization for data-intensive or computation-intensive tasks. 
A distributed environment is proposed for exploring correlations of large-scale simulation datasets. Well-designed data structures can be used to improve the performance of distributed visualization, like the PSH ~\cite{lefebvre2006perfect} and quadtree~\cite{xie2014visualizing}. GPU-based parallel computing can further improve performance~\cite{shih2016parallel}.
Likewise, the runtime efficiency of graph sampling~\cite{meidiana2019topology} and sparsification~\cite{arleo2017graphray} can be benefited from distributed computing. Note that these studies emphasize on the performance issues.
There is little effort made on data privacy in distributed visualization. 
Saha et al.~\cite{saha2017see} introduce decentralized data stochastic neighbor embedding (dSNE) to enable embedding and visualization of sensitive neuroimaging data. Similarly, a decentralized brain imaging data analysis is proposed with new data processing and visualization algorithms~\cite{plis2016coinstac}.


\subsection{Federated Learning}

The fundamental idea of federated learning (FL)~\cite{konevcny2016federated} is to learn an integrated model with data distributed on clients. It can balance performance and communication efficiency while preventing leakage of sensitive data. A large number of studies have been made on the usability, scalability, and performance of FL~\cite{li2019federated,wei2019multi}. For instance, an optimized federated learning strategy is proposed to train a high-quality centralized model~\cite{konevcny2016federatedoptimization}. By investigating enhanced privacy protection algorithms, different levels of privacy protection can be achieved in FL at a minor loss in model performance~\cite{mcmahan2017learning,geyer2017differentially,bonawitz2016practical}. A recent trend is to incorporate FL within a variety of application scenarios~\cite{huang2019patient,huang2018loadaboost,liu2020fedvision}. Concerning the distribution of training data, federal learning approaches can be classified into three categories:  horizontal federated learning (HFL), vertical federated learning (VFL), and Federated Transfer Learning (FTL)~\cite{yang2019federated}. In this paper, we focus on HFL, which can be applied in scenarios where datasets mastered by clients are collected respectively but share the same properties.
\section{Federated Visualization}

\begin{figure*}[t]
\centering

\includegraphics[width = \linewidth]{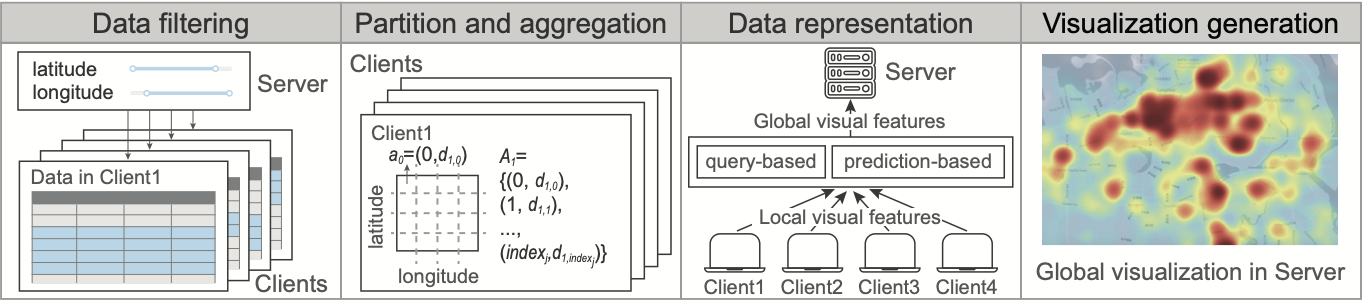}

\caption{
The pipeline for federated visualization consists of four main stages.
}
\label{fig2}
\end{figure*}

The goal of federated visualization is to protect privacy in a decentralized visualization framework. Suppose that a set of datasets $\textbf{D}=\{D_1, D_2, D_3, ..., D_N\}$, owned by clients $\textbf{C}=\{C_1, C_2, C_3, ..., C_N\}$, respectively. 
Assume that $V_i$ is the visualization of $D_i$, and $\textbf{V}$ is the visualization of $\textbf{D}$. Without loss of generality, $V_i$ and $\textbf{V}$ share the same visual transformations and mappings, resulting in a set of visual features, denoted as $\textbf{V}=\{VF_1, VF_2, ..., VF_M\}$. Here, $VF_i=\{vf_{i1}, vf
_{i2}, ..., vf_{iM}\}$ is a vector of visual features, such as position, color, size, and shape. $\textbf{V}$ represents the global visual features of $\textbf{D}$. 
In the conventional parallel visualization, $\textbf{D}$ is collected and processed globally in a server $\textbf{S}$, exposing a risk of privacy leakage.
In this paper, we only consider the scenario that the visual features of each client have the same dimensions, which refers to the situation of \textbf{HFL}~\cite{yang2019federated}. 

We follow the flowchart of \textit{horizontal federated learning} and propose a two-stage federation strategy. Datasets are stored and processed locally in each client $C_i$. $C_i$ only transmits encrypted parameters $P_i$ which represents visual features and cannot be used to recover the original data. $P_i$ acts as the messenger of the original data $D_i$ and $\textbf{V}$, to avoid direct exposure of the raw data. The definition and construction of $P_i$ varies with different federation schemes (see Section 3.2).

\subsection{Pipeline}

The pipeline consists of four main stages, as shown in Figure~\ref{fig2}:

\begin{itemize}
\item[1)] \textbf{Data filtering}. 
First, the data scope $E_i$ associated with $D_i$ is specified by analysts, e.g., data in a specific latitude and longitude range, or data in a specific time period. It can be defined uniformly for all clients, or be specified interactively on the visual interface during the exploration process.

\item[2)] \textbf{Data partitioning and aggregation}. 
For each client, its data is partitioned in small pieces uniformly. The resolution of the partition can be predefined or interactively specified subject to employed visual forms. For instance, creating a heatmap requires to divide the data over a 2D grid. Similarly, generating a histogram of traffic flow in one week needs to define the granularity along the time axis. 

Then, data aggregation is performed over each binned range of the partition. The aggregated values are mapped into visual features, forming a visual feature set $A_i$. Each element $a_j$ of $A_i$ is a key-value pair, namely, $a_j = (index_j, d_{i,index_j})$. Here, $index_j$ denotes the index, and \textit{d} is the feature values of $index_j$. 

This process is performed in each client, resulting in visual features with identical dimensions. 

\item[3)] \textbf{Data representation}. This process requires to encode and encrypt visual features in each client and utilize them in the server. For that, we design two implementations, which are described in Section 3.2.

\item[4)] \textbf{Visualization generation}. Based on the encrypted visual features from local clients, the server composes a set of global visual features and generates a visualization. Section 3.3 describes the ways of visualization creation associated with two representations introduced in Section 3.2. 

\end{itemize}

\subsection{Federated Representation}

In the third stage, it is very important to keep visual features in each client from being exposed during the transmission process. We solve this issue by transmitting encrypted parameters that represent visual features.
In particular, two implementations are designed: 

\begin{itemize}
    \item \textit{Query-based}, which encodes local visual features using secure aggregation techniques and decodes them in the sever;
    
    \item \textit{Prediction-based}, which trains a prediction model in each client through federated learning, and predicts the result in the server by means of all parameters of local prediction models.
\end{itemize}

\subsubsection{The query-based scheme}
We ensure that all clients possess pair-wise secure communication channels and use secure aggregation~\cite{bonawitz2016practical} for encryption.

\begin{figure}[ht]
\centering

\includegraphics[width = \linewidth]{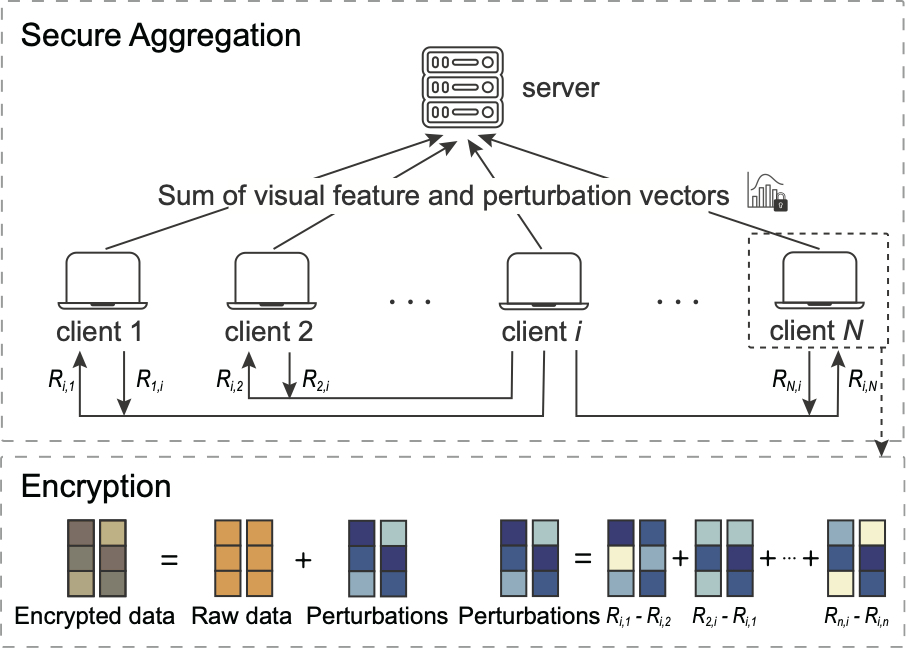}

\caption{
The flowchart of the \textit{query-based} scheme.
}
\label{qb}
\end{figure}

The \textit{query-based} scheme consists of three steps:

\begin{itemize}
    \item[1] Sampling random vectors. Each pair of clients first prepares a random vector for each other. That is, each client $C_i$ locally samples a random vector $R_{i,j}, j\in[1,N] \land i\neq j$ for each other client $C_j$. Specifically, $R_{i,j}$ and $R_{j,i}$ are a one-to-one pair. For example, $R_{1,3}$ and $R_{3,1}$ form the random vectors prepared by $C_1$ and $C_3$ for each other (Figure~\ref{qb}).
    \item[2] Exchanging random vectors and computing perturbations. Client $C_i$ and $C_j$ exchange $R_{i,j}$ and $R_{j,i}$ over their secure channels. To ensure transmission security, perturbations $P_{i,j}=R_{i,j}-R_{j,i},i\neq j$ are computed. The sum of visual feature and perturbation vectors is sent to the server: $D\_upload_i=VF_i+\sum_{j=1}^{N}P_{i,j}$.
    \item[3] Computing global visual features in the server. The server receives the perturbed vectors $D\_upload_i,i\in[1,N]$ uploaded by clients, and sums them:
    \begin{displaymath}
    \setlength{\abovedisplayskip}{3pt}
    \setlength{\belowdisplayskip}{3pt}
    D\_sum=\sum_{i=1}^{N}D\_upload_i =\sum_{i=1}^{N}VF_i+\sum_{i=1}^{N}\sum_{j=1}^{N}R_{i,j}-\sum_{i=1}^{N}\sum_{j=1}^{N}R_{j,i}=\sum_{i=1}^{N}D_i
    \end{displaymath}
    The result $D\_sum$ is guaranteed to be accurate visual features because the paired perturbations in $D\_upload_i$ are neutralized. And the values from each client will not be inferred.
\end{itemize}

\subsubsection{The prediction-based scheme}

Encoding visual features can be accomplished by precomputing a representation with the \textit{prediction-based} approaches. Conventional solutions include data fitting like logic regression, hashing, and neural network approaches. We choose to use neural network-based methods, and mimic the federated learning framework where all clients contribute to training a shared global model that represents global data features. Below we introduce our scheme based on neural network-based models. 

\begin{figure}[ht]
\centering

\includegraphics[width = \linewidth]{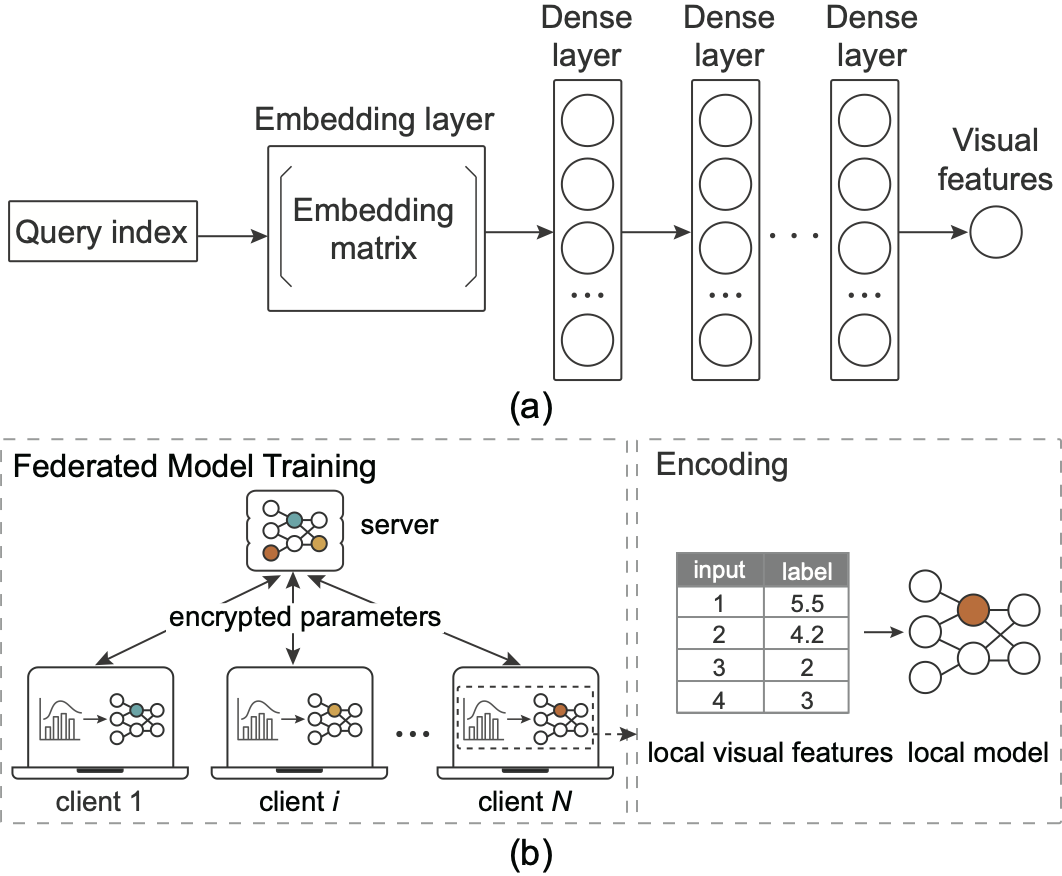}

\caption{
Illustration of the \textit{prediction-based} scheme. (a) The network structure of the employed NN model. (b) The training and prediction processes.
}
\label{fig3}
\end{figure}

We leverage a fully connected deep network~\cite{ramsundar2018tensorflow}, as shown in Figure~\ref{fig3} (a). 
It consists of an embedding layer that converts the input data into vectors to improve the efficiency of model training, and four fully connected layers. To enable the federation of the prediction models, we design an adapted NN model, which consists of a global part and local parts (See Figure~\ref{fig3} (b)). The server keeps the global part, while each client keeps a local part. We consider the visual feature set $A_i$ as the local model training set of $C_i$, where $index_j$ is used as the model input and $d_{i,index_j}$ is used as the label data, that is, the model output. We use the loss function defined as $\sum_{i}(d_{i,index_j}-\hat{d_{i,index_j}})^2$, where $i$ is the index of the underlying data piece, $\hat{d_{i,index_j}}$ denotes the value of the data predicted by the global model. We use the quadratic cost so that $d_{i,index_j}$ will be close to the average of $\hat{d_{i,index_j}}$ over all clients.

The training process consists of three steps:
\begin{itemize}
\item[1]  Initialization. The server S initializes a global neural network model $M\_fed$. Each client keeps a copy of the initialized model from the server. 
The server sends encrypted initial parameters of the global NN model to each client.

\item[2] Local training. Each client decodes the parameters sent from the server, and uses them to update the local models. Then, the local NN model  is iteratively trained for several rounds with $A_i$ as the training set, yielding updated parameters of the local NN model. The parameters are encrypted and sent back to the server.

\item[3] Federated averaging. The server performs a federated averaging~\cite{konevcny2016federated} over the set of encrypted parameters(in the form of $P_i(t)$, where $t$ represents one round, including the embedding matrix and weight matrix) from all clients, and computes updated global parameters, $P(t)$:
$P(t)=\frac{1}{N}\sum_{i=1}^{N} P_i(t)$. 
The parameters are then encrypted and sent to each client. 

The 2nd and 3rd steps are iteratively performed, until the loss function converges, or say, values of two consecutive iterations are adequately close. 
The trained NN model parameters are stored on the server for reuse.
\end{itemize}

\subsection{Federated Composition}

For a visual form to be created, relevant visual features are generated in each client. The encryption of model parameters or values is transmitted to the backend server. Visual features are reconstructed with respect to the employed scheme in the representation stage (Section 3.2).

\begin{figure}[ht]
\setlength{\belowcaptionskip}{-10pt}
\centering

\includegraphics[width = \linewidth]{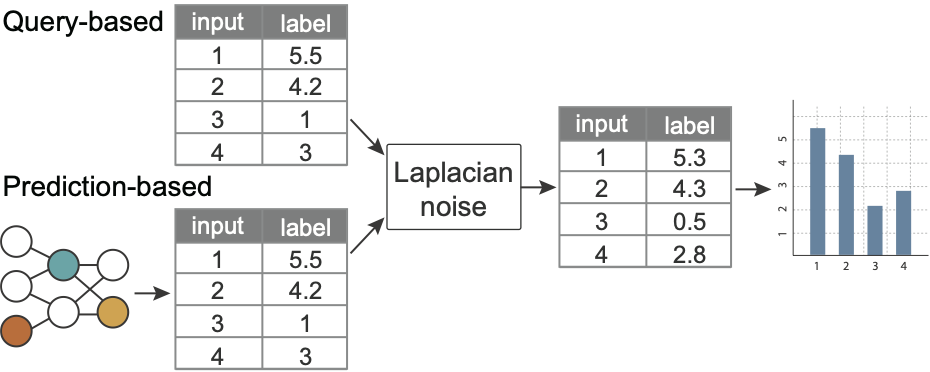}

\caption{
The process of federated composition with two schemes: the \textit{query-based} scheme and the \textit{prediction-based} scheme. In both schemes, Laplacian noise is added into the global visual features to prevent attackers from inferring individuals through global visual features. Smaller visual features will be added more noise.
}
\label{composition}
\end{figure}

\begin{itemize}

\item For the \textit{query-based} scheme, the server directly extracts accurate visual feature values.

\item For the \textit{prediction-based} scheme, the index $ index_j $ is used as the input of the trained global model $M\_fed$, and the output $ d_j $ can be expressed as:
\begin{displaymath}
\setlength{\abovedisplayskip}{3pt}
\setlength{\belowdisplayskip}{3pt}
d_j=\sum_{i=1}^{N}d_{i,index_j}/N
\end{displaymath}
It is approximately the average of aggregated values of each client within the data range $ index_j $. Therefore, visual features of the global data on $index_j$ are $N\times{d_{j}}$. In this way, we can get all visual features by feeding all indices into the global model $M\_fed$. 

\end{itemize}

In order to prevent the reconstructed global visual feature values from leaking privacy, we employ differential privacy~\cite{dwork2006calibrating} before visualization composition. When the visual feature value is small, it is easy to re-identify individuals~\cite{sweeney2002k}. Thus, we add more noises to visual features with relatively small values. As shown in Figure~\ref{composition}, after reconstructing the global visual features, we use the Laplace mechanism to add the random noise into the visual features. Then, the visual feature values with noise can be used for federated visualization to achieve privacy protection.

\section{Visual Interface}

We design and develop a prototype system for federated visualization.

\begin{figure*}[t]
  \centering
  
  \includegraphics[width = \linewidth]{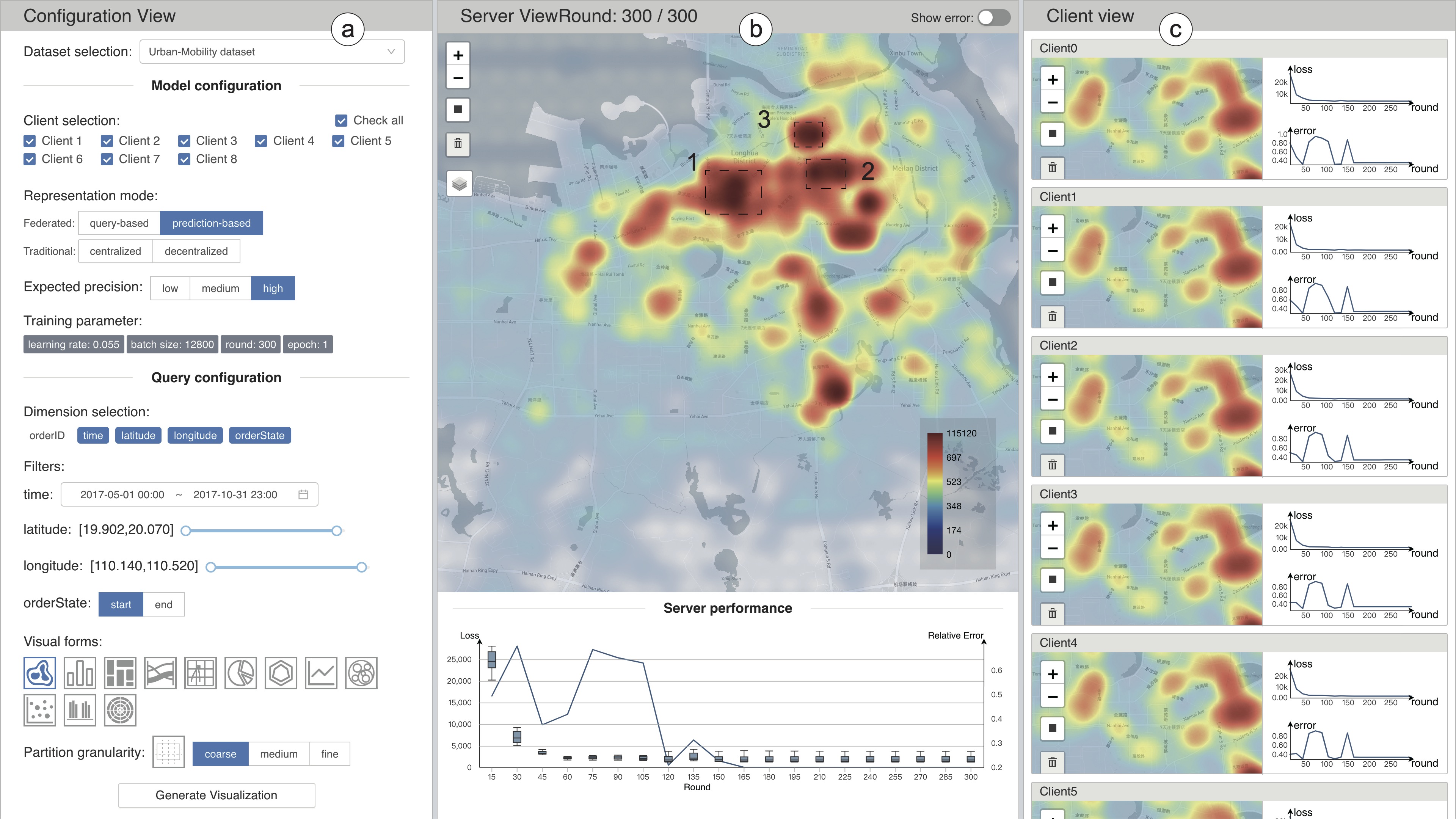}
  
  \caption{
    Demonstration of our approach with the Urban-Mobility dataset. Here, one server and eight clients are employed. The \textit{prediction-based} scheme is used to fulfill the federation of visualization. (a) The configuration view allows users to configure the relevant parameters of the federated scheme and query conditions. (b) The server view displays the global visualization result and the dynamic performance of the server. (c) The client view shows the local visualization and its performance.
  }
  \label{fig:teaser}
  \end{figure*}

\subsection{Interface}


The visual interface (Figure~\ref{fig:teaser}) has three main modules: the configuration view, server view and client view. 
The configuration view includes model and query configuration. 
In the model configuration module, the user specifies the clients participating in the federated visualization, and then selects the federated representation scheme to be adopted. If the \textit{prediction-based} scheme is specified, the user needs to further define the expected precision, which indirectly determines the model training parameters.
In the query configuration module, the user selects the dimension to explore and specifies the data range. Several commonly-used visual forms are listed. After selecting a visual form, the corresponding data partition will be displayed. The user needs to select the granularity of data partition. After completing all configuration items, the user can obtain the federated visualization results.
The server view shows the global visualization and the server performance. To observe the precision of the results, the user can view the error in the performance line chart, or switch to the error chart.
In order to make users understand the federated framework intuitively, the results and performance of each client using its own local data for visualization are displayed in the client view. Users can clearly compare local and global results.


\subsection{Federated chart generation}

Below, we show the creation of a variety of aggregation-based visualizations, ranging from structured tabular data to hierarchical datasets. Both the \textit{query-based} scheme and the \textit{prediction-based} scheme can be leveraged to get global visual feature values for visualization creation. The flexibility of federated visualization is highlighted in each example.

\textbf{Histogram}. To create a histogram, the number and height of bars need to be determined. Suppose the target histogram has 7 bars which represent 7 days of a week, and the height of each bar represents the traffic volume of a day. To generate it, the local data in each client is divided into 7 pieces, each of which the total traffic volume is computed. Thus, a local visual feature set for visual mapping is obtained in each client. To get the global visual features of the entire dataset, on one hand, we can use the \textit{query-based} scheme to directly encrypt and merge visual features of each client; on the other hand, the \textit{prediction-based} scheme can be used to fit the global visual features where the local visual feature set is used as the local training set. 
To further explore the finer granularity of the information, we create a stacked histogram. The underlying data of each bar is divided and computed to obtain a new visual feature set for visual mapping of the stacked histogram.

In general, the way of creating a histogram can be applied to generating other chart types, such as line chart, pie chart, violin plot, area chart, and radar chart.

\textbf{Heatmap}. The heatmap in Figure~\ref{fig:teaser} consists of 15960 (190$\times$84) grids divided by uniform longitude and latitude. Each grid cell has its index and value. The accumulated density in each grid cell is encoded by hue, which represents the number of records in the corresponding geographic area. On each client, a heatmap is calculated by local data. Those heatmaps, reflecting the distributions of dataset on each client, have the same cell index. Our goal is to get an overall heatmap which reflects the distribution of all clients' dataset without leaking privacy.

Similarly, to create an ODMap (Origin-Destination flow map), the local data is partitioned according to the latitude and longitude of the map. In each resultant part, the partition is performed along the latitude and longitude of the destinations. Then, the traffic flow from the origins to the destinations is computed, resulting in a four-dimensional division. When the division is relatively dense, the \textit{predictive-based} scheme is more time-consuming than the \textit{query-based} one, because considerable visual feature values need to be trained.

\textbf{Sankey diagram}. Sankey diagrams are a specific type of node-link diagrams. Several entities are represented by rectangles or text, and their links are represented with arrows or arcs whose widths are proportional to their flow quantities. To create a sankey diagram, flow quantities from different clients should be integrated without merging data. To solve this issue, each client defines data statistics rules according to the visual analysis task, divides and counts the local data according to the rules. For example, when analyzing population migration between different countries, each client divides and counts \textit{in} and \textit{out} fields based on different countries. These processed data are then encrypted and merged for final visualization.

\textbf{Squarified Treemaps}. Federated visualization is capable of creating recursive layouts, such as treemap, sunburst, and circle packing diagram. The data structure of hierarchical data (where the data is organized into a tree structure) is more complex than that of tabular data. Without loss of generality, the tree structures in all clients are assumed to be identical. For example, in the case of aggregating street sign-in information from various regions of the country from different clients, the hierarchy of the geographic locations of the clients is the same. Thus, to craft a treemap, each client only needs to count the numbers of corresponding leaf nodes and send their encryptions, then the server can summarize and render the treemap.

\subsection{Federated Visual Analysis}

Visual exploration with federated visualization forms a seamless composition of multiple client and server steps. First, the visualization configurations of federated visualization are determined in the front-end. Users can specify the participating, the representation mode, the data range, and the visual form in the configuration view. After the user selects the dimension and visual form, the system will automatically complete the visual mapping. Second, the configurations are sent to the server. The server determines the scheme based on the chosen representation mode and sends the configurations to all clients. Third, clients process the local data according to the representation mode and send the results back to the server. Fourth, the front-end gets visual feature values from the server and creates the visual forms. 

Several basic interactions, such as selection, navigation, and filtering are supported. When the user clicks the ``Generate Visualization'' button, the information transmission between the server and clients will be triggered. For the \textit{query-based} scheme, the dataset stored in each client is queried. For the \textit{prediction-based} scheme, the model needs to be re-trained, and related parameters are updated for each new query. Pre-computing results of all or parts of model configurations can eliminate the workload of model re-training in run-time visual queries, as has been employed in previous works~\cite{mei2019rsatree}.
\section{Evaluation}

We implement a web-based prototype system. The source code of our system is available at: \url{https://github.com/ZeroWangZY/fed-vis}. The back-end is written with Python. The front-end visualization is implemented with a combination of HTML5, JavaScript, and D3.js. SocketIO is used to communicate between clients and the server. TensorFlow is used to implement our deep learning model. 
All clients are simulated on a single machine with eight simulation nodes, which has an i7-8700 processor and a GTX 1660Ti GPU.

\subsection{Quantitative evaluation}

The quantitative evaluation compares our two schemes with the centralized counterpart with three measures: performance, accuracy and visual quality.


\begin{table*}[t]
   \setlength{\abovecaptionskip}{0pt}
  \setlength{\belowcaptionskip}{-8pt}
  \vspace{-2mm}
  \centering
  \caption{Quantitative results of the time measurement for heatmap, histogram, i.i.d. treemap and non-i.i.d. treemap using the \textit{prediction-based} scheme and the \textit{query-based} scheme under different settings of the number of client and training round.}
    \begin{tabular}{ c | c|c c c | c c c | c c c }
     \hline
     \multicolumn{2}{c|}{\#client} & \multicolumn{3}{c|}{3} & \multicolumn{3}{c|}{5} & \multicolumn{3}{c}{8}\\
     \hline
     \multicolumn{2}{c|}{\#round} & 50 & 150 & 300 & 50 & 150 & 300 & 50 & 150 & 300\\
                                     
    \hline
     \multirow{3}{*}{Prediction-based} & Heatmap($380\times168$) & 17.5633 & 52.3755 & 102.8733 & 29.8091 & 87.8196 & 174.8193 & 47.6048 & 140.334 & 278.6307 \\
                                & Histogram($24\times7$) & 1.8203 & 4.7769 & 7.6263 & 2.7338 & 6.6951 & 12.6123 & 3.8703 & 10.2641 & 19.6864 \\
                                & Treemap(6000) & 2.7662 & 6.433 & 12.7318 & 4.4036 & 11.8682 & 20.1864 & 6.3223 & 16.3948 & 31.6249 \\
    \hline
     \multirow{3}{*}{Query-based} & Heatmap($380\times168$) & \multicolumn{3}{c|}{2.2464} & \multicolumn{3}{c|}{2.4535} & \multicolumn{3}{c}{2.7742} \\
                    & Histogram($24\times7$) & \multicolumn{3}{c|}{2.0591} & \multicolumn{3}{c|}{2.106} & \multicolumn{3}{c}{2.1822} \\
                    & Treemap(6000) & \multicolumn{3}{c|}{2.21} & \multicolumn{3}{c|}{2.1269} & \multicolumn{3}{c}{2.1946} \\
    \hline
    \end{tabular}
  \label{tab:1}
\end{table*}

\begin{table*}[t]
  \setlength{\abovecaptionskip}{0pt}
  \setlength{\belowcaptionskip}{-8pt}
  \vspace{-2mm}
  \centering
  \caption{Quantitative results of the accuracy measurement for heatmap, histogram, i.i.d. treemap and non-i.i.d. treemap using the \textit{prediction-based} scheme under different settings of the number of client and training round.}
    \begin{tabular}{ c | c|c c c | c c c | c c c }
     \hline
     \multicolumn{2}{c|}{\#client} & \multicolumn{3}{c|}{3} & \multicolumn{3}{c|}{5} & \multicolumn{3}{c}{8}\\
     \hline
     \multicolumn{2}{c|}{\#round} & 50 & 150 & 300 & 50 & 150 & 300 & 50 & 150 & 300\\

    \hline
     \multirow{3}{*}{Prediction-based} & Heatmap($380\times168$) & 0.1286 & 0.0804 & 0.0821 & 0.1584 & 0.1185 & 0.1176 & 0.1783 & 0.1587 & 0.141 \\
                        & Histogram($24\times7$) & 0.1111 & 0.0261 & 0.0233 & 0.041 & 0.0307 & 0.0296 & 0.0458 & 0.0324 & 0.0286 \\
                        & Treemap i.i.d.(6000) & 0.0146 & 0.001 & 0.002 & 0.0025 & 0.0025 & 0.0008 & 0.0023 & 0.0025 & 0.0026 \\
                        & Treemap non-i.i.d.(6000)  & 0.0109 & 0.0003 & 0.0005 & 0.001 & 0.0007 & 0.0006 & 0.0005 & 0.0009 & 0.0012 \\
    \hline
     \multirow{1}{*}{Query-based} & * & \multicolumn{3}{c|}{0} & \multicolumn{3}{c|}{0} & \multicolumn{3}{c}{0} \\
    \hline
    \end{tabular}
  \label{tab:2}
\end{table*}


\subsubsection{Performance}
The response time includes time spent on data filtering, data partition and aggregation, data representation and visualization generation. The difference between federated visualization and traditional visualization methods lies in the data representation phase. The timing measurement indicates the impact of factors such as the granularity of data partitioning, the number of training rounds, and the number of clients. 


\textbf{Query-based scheme.}
Compared with traditional centralized visualization, extra time is spent on data encryption. It mainly includes three parts: generating random vectors, exchanging random vectors, encrypting and decrypting data. We assume that both the number of clients and the granularity of data partitioning affect the response time.

\textbf{\textit{Number of clients:}} The number of clients is set to 3, 5, 8. The response time presents a positive growth trend with the increasing of the number of clients (Table~\ref{tab:1}). This result means that the response time increases when the number of participating clients gradually increases. 
It is clear that the increasing rate of the response time depends on the underlying encryption mechanism.
 Exchanging random vectors dominates the computing time. This is because its time complexity is $O(n^2)$.

\textbf{\textit{Granularity of data partitioning:}}
Different chart types have different granularities, that is, the number of visual features, yielding varied response times (Table~\ref{tab:1}). The response time of heatmap is the longest and that of treemap is longer than that of histogram. This is because their granularity of data partitioning decreases and is $380\times168$, 6000, $24\times7$, respectively.

\textbf{Prediction-based scheme.}
The time overhead for the \textit{prediction-based} scheme lies in model training. Several factors may influence the time of model training, such as the granularity of data partitioning, the training settings, the number of clients, and the data distribution.

\textbf{\textit{Granularity of data partitioning:}}
Table~\ref{tab:1} shows the relationship between the response time and the granularity of data partitioning. The result is similar to that of the query-based scheme, indicating that when the number of visual features to be fit increases, the training time increases too.
Compared with the result of the query-based scheme in Table~\ref{tab:1}, the \textit{prediction-based} scheme has a larger time consumption. 

\textbf{\textit{Training settings:}}
The response time presents a positive growth trend as the number of training round increases (Table~\ref{tab:1}). Meanwhile, the response time of charts with a finer granularity will be relatively longer to fit more visual feature values.

\textbf{\textit{Number of clients:}}
The response time increases proportionally with the number of clients (Table~\ref{tab:1}). As the number of participating clients increases, the amount of training data increases, more time is needed to reach a model convergence.


\begin{figure*}[t]
\setlength{\abovecaptionskip}{0pt}
\setlength{\belowcaptionskip}{-4pt}
\centering
\includegraphics[width = 0.8 \textwidth]{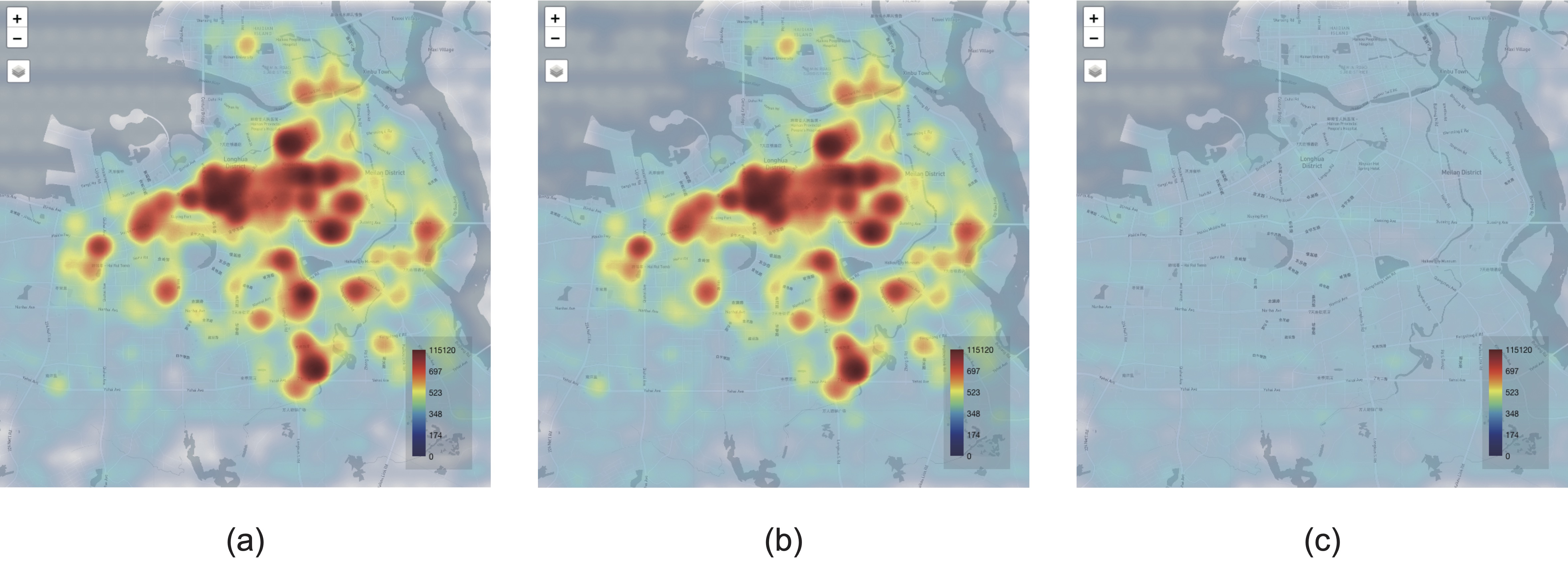}
\caption{Applying the \textit{prediction-based} scheme in querying a heatmap. From left to right: (a) the predicted result, (b) the exact result, (c) the difference map. To make the difference visible,  the difference values are enlarged by 50 times.}
\label{fig-pre-error}
\end{figure*}

\subsubsection{Accuracy}

With the \textit{query-based} scheme, 
a small amount of Laplacian noise has been added to the queried visual features, and the results are still relatively accurate. In contrast, errors occur with the \textit{prediction-based} scheme because visual features are approximately generated. Several factors, such as the granularity of data partitioning, the training settings, the number of clients, and the training data distribution, may influence the accuracy of the visualization. 
A accuracy measurement is considered. We use relative error (RE) metric to evaluate the difference between two visualizations. Suppose that a visualization contains $N$ feature values. Federated visualization fits original $N$ visual features $\hat{y_1}$, ..., $\hat{y_N}$, and returns $N$ values $y_1$, ..., $y_N$. The RE of the visualization is defined as:

\begin{displaymath}
\setlength{\abovedisplayskip}{3pt}
\setlength{\belowdisplayskip}{3pt}
RE
=\frac{\sum_{i=1}^{N} {\left| y_i-\hat{y_i} \right|}}{\sum_{i=1}^{N} \left| y_i \right|}
\end{displaymath}

\textbf{\textit{Training settings:}}
RE is inversely proportionally to the number of training rounds (Table~\ref{tab:2}). This means that the global model gradually reaches the overall optimal when the number of training rounds increases. 

\textbf{\textit{Number of clients:}}
As shown in Table~\ref{tab:2}, RE does not vary too much with the number of clients.

\textbf{\textit{Granularity of data partitioning:}}
Table~\ref{tab:2} shows that there is no relationship between accuracy and the chart type, that is, the granularity of data partitioning. The result of the heatmap is relatively larger than the other two charts. This may be because the distribution of a heatmap is uneven, and there are many visual features whose value is 0.

\textbf{\textit{Training data distribution:}}
Many works~\cite{zhao2018federated} have studied the distribution of training data in the field of federated learning. The training data on each client can be independently (i.i.d.) identically distribution or non-i.i.d.. 
In practice, we randomly generate two pieces of treemap data, which are i.i.d. and non-i.i.d., to simulate actual usage scenarios and analyze the impact of data distribution on the response time.
As shown in the Table~\ref{tab:2}, there is almost no difference in response time between two distributions, indicating that the data distribution has little effect on the accuracy with our method.



\subsubsection{Visual quality}

We compare our results with those generated by conventional centralized visualization, and generate an error chart. For example, we generate a difference map (Figure~\ref{fig-pre-error}) for heatmap. We use the same colormap as heatmap to encode the absolute value of the difference. The relative error of the obtained result in Figure~\ref{fig-pre-error} (a) is 0.003. This indicates little difference between the result with the \textit{prediction-based} scheme and the exact result. Two results are visually identical.

\subsection{Case studies}

In this section, we describe how our framework facilitates users to achieve aggregated visual queries and use various visual forms for visual analysis while protecting data privacy. We invite three users to participate in case studies on three real-world datasets. Each dataset's records are distributed over eight clients and are non-i.i.d. across clients (i.e., they are not independent and identically distributed (i.i.d.)). The system provides four modes: traditional centralized mode, traditional decentralized mode, federated \textit{query-based} mode and federated \textit{prediction-based} mode. Users can switch modes according to analysis requirements.

\subsubsection{Case One: The United States Cancer Statistics dataset}

The United States Cancer Statistics (USCS) dataset is the official statistics on cancer incidence by region, race, sex, year, and leading cancer site. We extracted cancer incidence data from 2007 to 2017, including 222133 data records.

\begin{figure}[ht]
\centering
\includegraphics[width =  \linewidth]{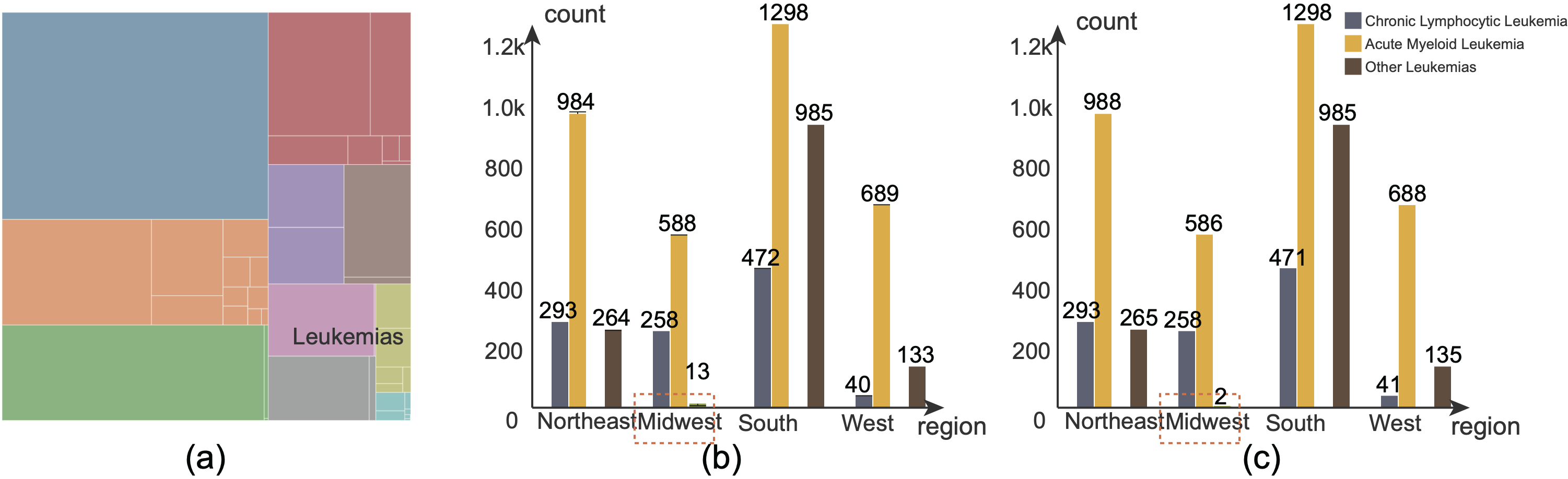}
\caption{The treemap of traditional desentralized mode shows that the incidence of Leukemias is relatively small (a). The histograms of the federated \textit{query-based} mode (b) and the traditional decentralized mode (c) have a little difference. The error bars are shown in (b). }
\label{case1}
\end{figure}

First, the user chooses the traditional decentralized mode in the configuration view. He uses a treemap to visualize the dimension ``Cancer site'' and observes the incidence of different cancer sites (Figure~\ref{case1} (a)). He finds that the incidence of Leukemias is relatively small. Then he queries the incidence of Leukemias subcategories in different regions and filtered by ``sex = male'', ``race = Black or African American'' and ``year = 2010''. The results are presented in a user-specified grouped histogram, as shown in Figure~\ref{case1} (c). He notices that there are only two men in the Midwest region who have Chronic Myeloid Leukemia cancer. Such a pattern is sensitive and has a high privacy risk. The attacker is likely to infer sensitive attributes or reveal individual identities with knowledge of quasi-identifiers.

Then, he chooses the federated query-based mode to conduct the same exploration (Figure~\ref{case1} (b)). He finds that even if he uses the same query conditions, results are slightly different. This makes it difficult to re-identify raw data items. When he switches to the error chart, he observes that when the visual feature value is small, the error value is relatively large, which increases the difficulty for the attacker to re-identify.
This shows that our scheme can protect privacy well.

\subsubsection{Case Two: The MovieLens dataset}

This dataset describes ratings from MovieLens, a movie recommendation service. It contains 1000208 records, including movie, user, and rating information. 
In order to simplify the case, we evenly distribute the rating data of each movie to multiple clients so that the weight can be ignored when calculating the average rating.

\begin{figure}[ht]
\centering
\includegraphics[width =  \linewidth]{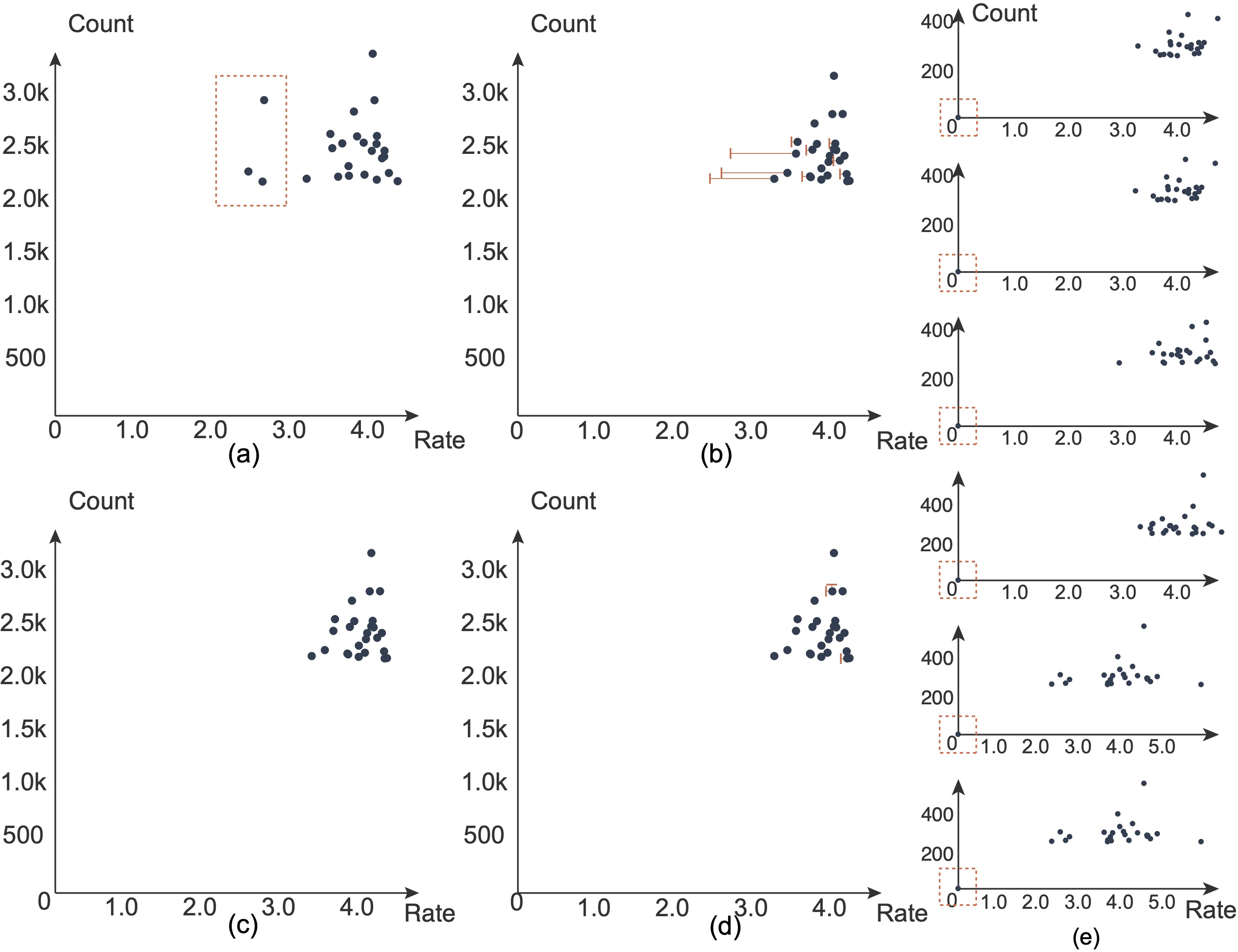}
\caption{The visualization result of the \textit{query-based} scheme (a) and the \textit{predition-based} scheme (c). The error chart of the the \textit{query-based} scheme (b) and the \textit{predition-based} scheme (d). There are many client views which contain dots at (0,0), as marked by red dashed boxes in (e).}
\label{case2}
\end{figure}

First, the user chooses the \textit{query-based} mode and analyzes the ratings of movies of interest using a scatterplot.
The scatterplot integrates the rating and number of rating records of all clients. Each point in the scatterplot represents a movie. Under our federated visualization framework, he can see more comprehensive patterns while protecting privacy.

However, when he observes the ratings of the movies ``Godfather, The '' (1972), ``Star Wars: Episode IV - A New Hope'' (1977), ``Titanic'' (1997), he notices that the RE presented in the server performance panel is 0.015, which is relatively large compared with other results. Then he switches to the error chart and observes that the errors of these three movies are relatively large, as shown in Figure~\ref{case2} (b). 
In the client view, he notices several clients have points at (0, 0), which means that the client does not have corresponding movie rating data (Figure~\ref{case2} (e)). 
Specifically, he finds that there is no rating data for ``Star Wars: Episode IV-A New Hope'' (1977) in the client1, client2 and client5, and there is no rating data for ``Godfather, The'' (1972) and ``Titanic'' (1997) in the client6, client7, and client8. With the \textit{query-based} scheme, each client needs to upload visual feature vectors of the same length. Each item in the vector corresponds to the value of a visual feature index. If there is no data corresponding to a certain index locally, the value will be automatically filled with 0. In this case, when there is no movie rating data, that is, there is no data corresponding to this index locally, its value is processed as 0. This causes the movie rating to be pulled down in the global visualization. In addition, for this kind of missing data situation, the \textit{query-based} scheme is not always incorrect. If the visual feature is obtained by counting items, it makes sense to treat no data record for a certain visual feature index as 0. This means that there is no data record that meets the query conditions.

The \textit{prediction-based} scheme can be well compatible with this situation. Figure~\ref{case2} (c) shows the result of the same aggregated visual query, with relatively small errors. This is because the \textit{prediction-based} scheme does not handle missing data items. 
For example, in this case, to fit the average rating of ``Star Wars: Episode IV-A New Hope'' (1977), only the data in the 
client3, client4, client6, client7, client8 are involved in the training to ensure the accuracy of the results. 
The \textit{prediction-based} scheme requires federated model training. As the training round increases, the RE of the visualization results decreases. The training process makes the response time of the \textit{prediction-based} scheme longer than that of the \textit{query-based} scheme. In short, if the user's analysis scenario is complex, which requires better compatibility and does not require fast response time, the \textit{prediction-based} scheme can be used. If the user needs a fast response time, he can choose the \textit{query-based} scheme.

\subsubsection{Case Three: Urban data exploration}
In the above two cases, we use the federated visualization framework to create treemaps, grouped histograms, and scatterplots to complete the privacy-preserving visual analysis. We can create more different charts to visualize the results of aggregated visual queries. This case uses the Urban-Mobility dataset, which contains 8,283,605 records, and is obtained from Didi Chuxing Technology Co., the biggest ride-hailing service company in China. The basic unit of recorded data is a taxi order, which contains information about the latitude and longitude of order start and order end, the time of order start and order end. The time spans from May 1st, 2017, to October 31st, 2017. In this case, the user can choose either the \textit{query-based} or the \textit{prediction-based} scheme according to their requirements.

\begin{figure}[ht]
\centering
\includegraphics[width =  \linewidth]{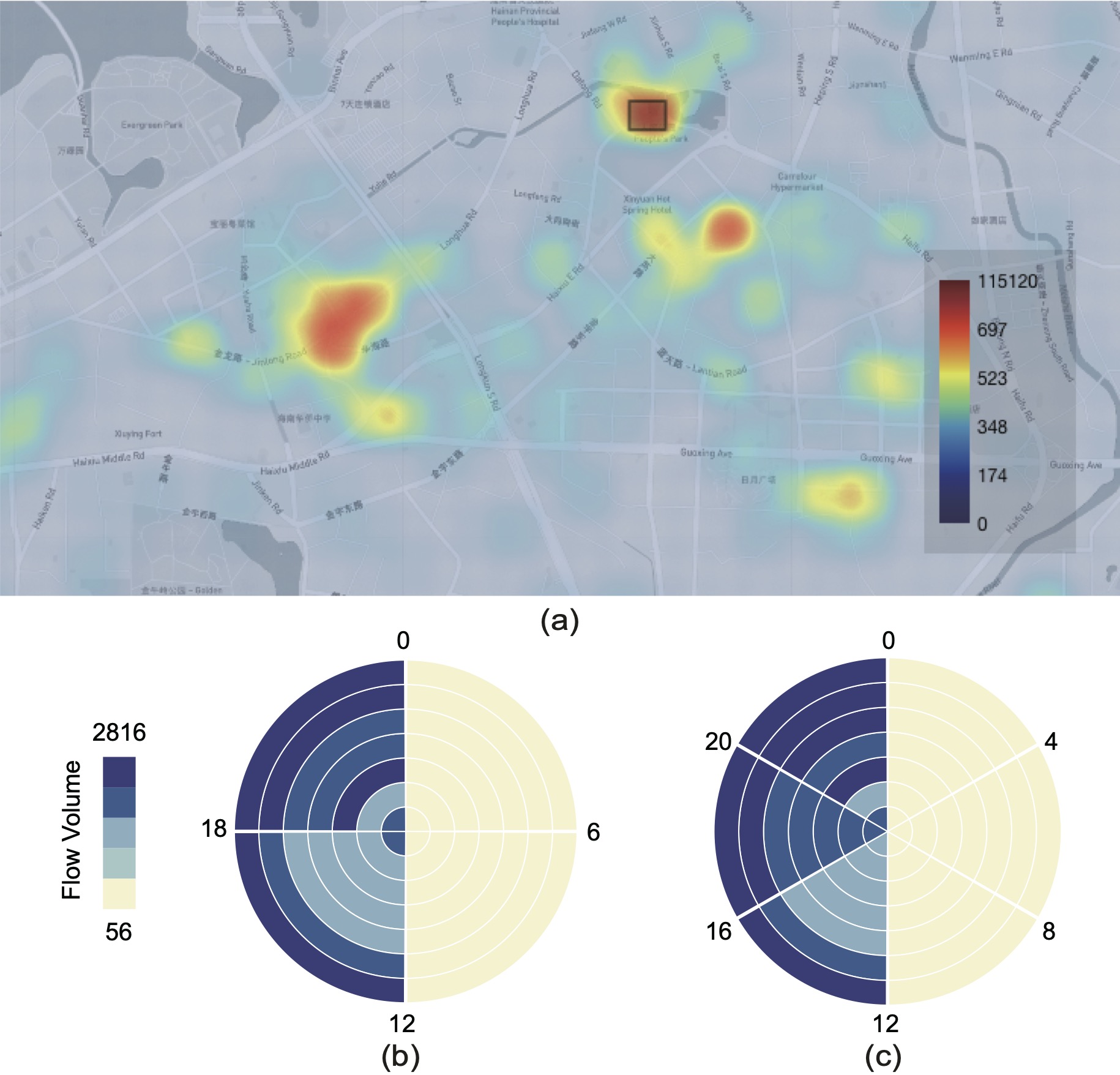}
\caption{Locations near the night fair have relatively heavy traffic
between 18 pm and 24 pm. Adjusting the hours per arc to 4 further
indicates that the traffic between 20 pm and 24 pm is heavy. The color of bars encodes different traffic flow volumes.}
\label{night-fair-total}
\end{figure}

First, the user chooses the \textit{prediction-based} scheme. He uses a heatmap to analyze the traffic flow from May 1st to October 31st. Figure~\ref{fig:teaser} indicates several locations of high traffic intensity. To further study the constant high intensity pattern, three locations in Figure~\ref{fig:teaser} are explored. Location 1 presents the highest intensity because it contains several shopping malls and apartment districts. The same reason holds for location 2. In location 3, there is one of the most famous night fairs in the city. To validate this pattern, a polar heatmap is created using the same dimensions (Figure~\ref{night-fair-total}). Figure~\ref{night-fair-total} (b) indicates that most traffic happens between 18 pm to 24 pm every day. After modifying the granularity of data partition, a more detailed aggregation shown in Figure~\ref{night-fair-total} (c) shows that traffic between 20 pm and 24 pm have a large proportion in daily traffic. 

\begin{figure}[ht]
\centering
\includegraphics[width =  \linewidth]{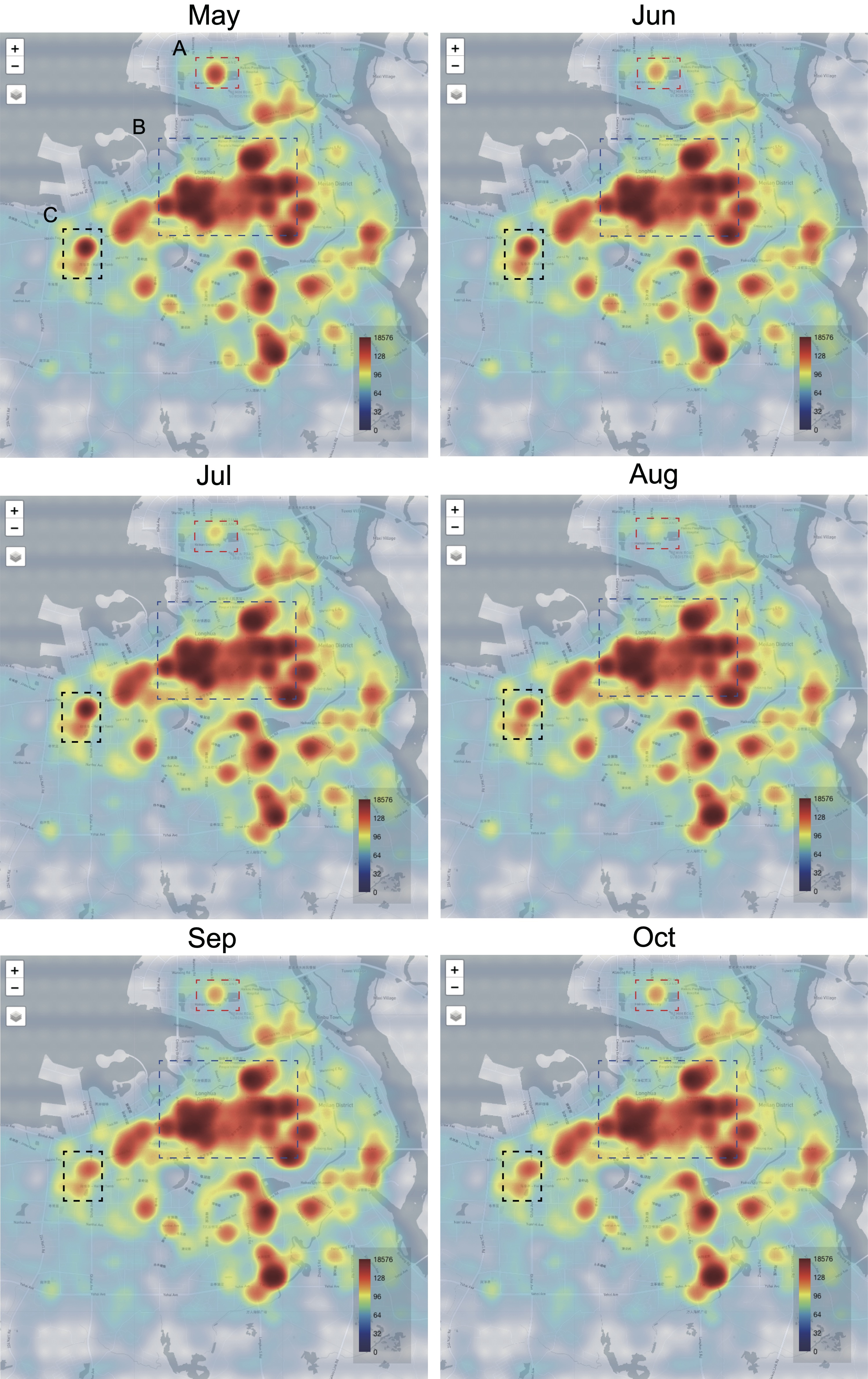}
\caption{Heatmaps show visual query results from May to October.
Three locations exhibit different spatio-temporal patterns. }
\label{heatmap-by-month}
\end{figure}

To study spatio-temporal patterns, Six queries are made to identify locations of intensive flows, resulting in six heatmaps 
(Figure~\ref{heatmap-by-month}). 
These heatmaps disclose locations of three types of temporal mobility patterns: the intensity first increases and then decreases (location C); the intensity first decreases and then increases (location A); the intensity stays stable (location B). By zooming in the view of location A around a university, we can see that its traffic intensity decreases in July and August and increases in September and October. This is because most students leave the university during the summer vacation. There are two specific points in location C: a long-distance bus station and a metro exit next to a hospital. The bus station and hospital are the main reason for the high traffic peak in the summer vacation.

\begin{figure}[ht]
\centering
\includegraphics[width = \linewidth]{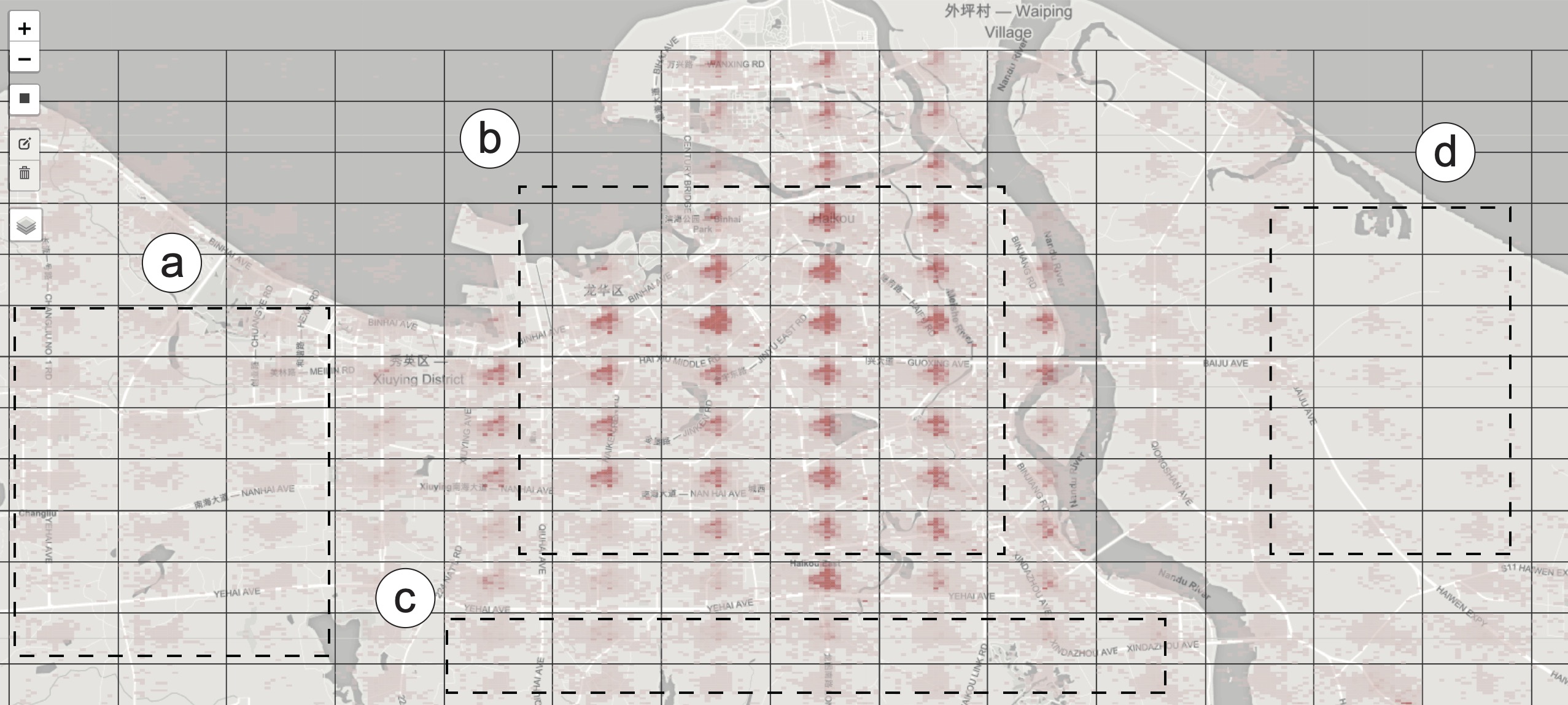}
\caption{Four districts with different flow patterns. }
\label{odmap}
\end{figure}

To further study the mobility modes over regions, ODMap~\cite{wood2010visualisation} is employed to show the links among sources and destinations. It is essentially a form of origin in which rows and columns represent the locations of flow origins and the locations of destinations, respectively. Figure~\ref{odmap} indicates that there are different flow patterns between districts. Most grids in the center of ODMap present a similar distribution and high intensity, while grids of the left part, bottom part, and right part exhibit distinctive distributions. These lead to two observations: 4 different districts present four different mobility modes; The downtown region has the highest intensity in both the heatmap and ODMap.

In short, the federated visualization framework can support the privacy-preserving visualization of the aggregated visual query.

\subsection{Expert Interview}

To evaluate the effectiveness of federated visualization, we conducted one-on-one interviews with three experts.

\textbf{Background.} All experts have experiences in visual analysis. In a prior discussion, they agreed that privacy is a big concern in visually analyzing decentralized data. A comprehensive analysis of a problem is not possible if data can not be shared. They have no knowledge about federated visualization.

\textbf{Process.} In each interview, we first introduced the background of federated visualization, our visual interface and chart examples in gallery, and demonstrated how the system works with the first case. Then we asked them to freely explore three datasets, answered their questions and observed their behaviors. Finally, we collected their feedback. This process took approximately 60 minutes.

\textbf{Feedback.} Overall, all experts felt that our system has no difference from conventional visual analysis systems and can be easily analyzed. They confirmed this convenience to various chart types and some basic interactions. 
An expert liked the provision of different accuracies in the system, which makes it flexible for different privacy scenarios. 
Surely, a low accuracy has little influence on visual distributions but can shorten the system response time. An expert tried to re-identify an arbitrary individual, but failed. This verifies the effectiveness of our system.

All experts commented, ``\textit{It takes a long time to refresh the heatmap and ODMap interactively, while the polar heatmap can be refreshed in real-time.}'' To handle complex charts and large datasets, schemes for optimizing query performance like the pre-computation scheme, could be incorporated in the future. 
An expert reported, ``\textit{I cannot feel the results are inaccurate, and I am not confident in the results of the analysis.}'' And he further suggested us to present the inaccuracy of the result. For example, visual encodings of uncertainty can be strengthened to clarify the inaccuracy.
They also discussed the possibility of using federated visualization to create complicated visualizations. They hoped that our approach could be extended to support complex scenarios, such as graph data or multi-source datasets from different domains. 



\section{Discussion}

Below, we discuss our work in terms of generality, privacy-preserving and performance. We also summarize the limitations and suggest directions for further work.

\textbf{Generality.} Case studies on several visual forms verify the effectiveness of our approach. Actually, our approach is applicable to all visualization charts that can be generated through data aggregation (mean and addition). Essentially, our approach aggregates visual features from each client without obtaining the specific raw data of each client. This means that if visual features of a chart can be obtained by summing data attributes, privacy-preserving visualization is feasible. For example, creating a pie chart needs to count the amount of data items for corresponding sectors from clients.

The \textit{prediction-based} scheme has a better generality than the \textit{query-based} scheme. In particular, the \textit{query-based} scheme may yield less accuracy when there are missed attribute values. For example, the server needs to use data from three clients ($C_1$,$C_2$,$C_3$) to visualize the probability of two events $E_A$ and $E_B$. Both the data held by $C_1$ and $C_2$ can be used to calculate the probability of two events, while $C_3$ can only obtain the probability of $E_A$ and does not know the probability of $E_B$. In this case, the \textit{query-based} scheme with which each client needs to send all visual features does not work. If the probability of $E_B$ is set to be zero by $C_3$ to complement all visual features to participate in federated visualization, the overall accuracy decreases, because the client does not know the probability of $E_B$. In contrast, the \textit{prediction-based} scheme works in this scenario. If some attribute values are missing in a client, the client only employs existing data for local training. The missing data does not affect accuracy.



\textbf{Privacy preservation.}
Secure aggregation~\cite{bonawitz2016practical} is widely used in federated learning to encrypt the model gradients of each client and avoid backtracking the original data.
Secure aggregation can guarantee privacy under the threat conditions of \textit{server is honest-but-curious} and \textit{server can lie to client}, that is, all clients follow the protocol honestly, but the server may try to learn additional information in different ways. However, collusion among clients can result in  privacy leakage. Our approach employs secure aggregation to protect the privacy of transported parameters. In our implementation, if four of the five clients collude with each other, these four clients send invalid information to the server, such as zero. Consequently, the composed visualization only shows the data of the innocent client, which may lead to privacy exposure.

Conventional approaches seek to protect privacy through data anonymity. However, simply anonymizing data would lead to a low utility. The visualization composed from anonymized data might exhibit incorrect patterns and mislead the decision. Our approach aggregates data to obtain visual features, and allows for secured visual analysis within the federated visualization framework. As such, global visual features are relatively accurate.

\textbf{Performance.} Several factors, such as response time and accuracy, may influence the performance of federated visualization.
Compared with traditional centralized visualization, our approach has extra time overhead to achieve privacy protection.
The \textit{query-based} scheme integrates encrypted visual features on the server side, and computes the sum of visual features of each client without loss. Data encrypted transmission costs extra response time.
The \textit{prediction-based} scheme uses the data on each client to train a global model to fit global visual features. Ideally, its accuracy can reach 100\%. However, in real scenarios, models are often unable to fit completely, because of time constraints, data distributions (non-i.i.d. or i.i.d.), or federated learning characteristics.
Quantitative results show that the time increases with the accuracy. A balance between the accuracy and the time can be achieved.
In addition, as the amount of data required to create visual charts increases, obtaining more accurate results is more time-consuming. There is much capacity for further performance optimization. For example, the level-of-details scheme can be employed in visualizing a heatmap.

The \textit{query-based} scheme can generate more accurate results than the \textit{prediction-based} scheme, and has a better running performance. Thus, it is more feasible for time-critical situations. 

\textbf{Limitations and future work.} First, the number of clients should be larger than 3. Otherwise, the collected information in the server can be used to infer secured information. Second, clients should make a strict protocol agreement and be honest because our approach can not identify the collusion among clients. Third, our \textit{prediction-based} scheme re-trains the model after each user interaction. When the number of visual features to be trained reaches tens of thousands, the response time becomes even longer, resulting in unfriendly user experience. In the future, we can integrate more schemes into our federated visualization framework, not limited to aggregated visual queries.


\section{Conclusion}
This paper addresses an important aspect of decentralized visualization: privacy. The fundamental idea is to mimic the process of federated learning, and reformulate the visualization process with a new federated model. We propose two implementations: a \textit{query-based} scheme, and a \textit{prediction-based} scheme.

The \textit{query-based} scheme directly encrypts accurate results, resulting in relative short response time than the \textit{prediction-based} scheme.
The \textit{prediction-based} scheme can be inefficient when the parameter number of the prediction model is large, i.e., more than  ten thousands. We plan to reduce its computational complexity. Our work is the first attempt to tackle data privacy issues in a decentralized visualization framework. We hope that this work will inspire other researchers to further study the privacy preservation solution in distributed environment, which should be a meaningful direction for sensitive data collaborative analysis.

\ifCLASSOPTIONcaptionsoff
  \newpage
\fi



%

\bibliographystyle{IEEEtran}
\bibliography{bare_jrnl_compsoc}

\begin{thebibliography}{10}
\providecommand{\url}[1]{#1}
\csname url@samestyle\endcsname
\providecommand{\newblock}{\relax}
\providecommand{\bibinfo}[2]{#2}
\providecommand{\BIBentrySTDinterwordspacing}{\spaceskip=0pt\relax}
\providecommand{\BIBentryALTinterwordstretchfactor}{4}
\providecommand{\BIBentryALTinterwordspacing}{\spaceskip=\fontdimen2\font plus
\BIBentryALTinterwordstretchfactor\fontdimen3\font minus
  \fontdimen4\font\relax}
\providecommand{\BIBforeignlanguage}[2]{{%
\expandafter\ifx\csname l@#1\endcsname\relax
\typeout{** WARNING: IEEEtran.bst: No hyphenation pattern has been}%
\typeout{** loaded for the language `#1'. Using the pattern for}%
\typeout{** the default language instead.}%
\else
\language=\csname l@#1\endcsname
\fi
#2}}
\providecommand{\BIBdecl}{\relax}
\BIBdecl

\bibitem{plis2016coinstac}
S.~M. Plis, A.~D. Sarwate, D.~Wood, C.~Dieringer, D.~Landis, C.~Reed, S.~R.
  Panta, J.~A. Turner, J.~M. Shoemaker, K.~W. Carter \emph{et~al.},
  ``{COINSTAC: A Privacy Enabled Model and Prototype for Leveraging and
  Processing Decentralized Brain Imaging Data},'' \emph{Frontiers in
  Neuroscience}, vol.~10, p. 365, 2016.

\bibitem{saha2017see}
D.~K. Saha, V.~D. Calhoun, S.~R. Panta, and S.~M. Plis, ``{See Without Looking:
  Joint Visualization of Sensitive Multi-site Datasets},'' in \emph{Proceedings
  of the Twenty-Sixth International Joint Conference on Artificial
  Intelligence}, 2017, pp. 2672--2678.

\bibitem{oksanen2015methods}
J.~Oksanen, C.~Bergman, J.~Sainio, and J.~Westerholm, ``{Methods for deriving
  and calibrating privacy-preserving heat maps from mobile sports tracking
  application data},'' \emph{Journal of Transport Geography}, vol.~48, pp.
  135--144, 2015.

\bibitem{wang2018user}
X.~Wang, T.~Gu, X.~Luo, X.~Cai, T.~Lao, W.~Chen, Y.~Wu, J.~Yu, and W.~Chen,
  ``{A User Study on the Capability of Three Geo-based Features in Analyzing
  and Locating Trajectories},'' \emph{IEEE Transactions on Intelligent
  Transportation Systems}, vol.~20, no.~9, pp. 3375--3385, 2018.

\bibitem{chen2017vaud}
W.~Chen, Z.~Huang, F.~Wu, M.~Zhu, H.~Guan, and R.~Maciejewski, ``{VAUD: A
  Visual Analysis Approach for Exploring Spatio-Temporal Urban Data},''
  \emph{IEEE Transactions on Visualization and Computer Graphics}, vol.~24,
  no.~9, pp. 2636--2648, 2017.

\bibitem{zhou2018visual}
Z.~Zhou, L.~Meng, C.~Tang, Y.~Zhao, Z.~Guo, M.~Hu, and W.~Chen, ``{Visual
  Abstraction of Large Scale Geospatial Origin-Destination Movement Data},''
  \emph{IEEE Transactions on Visualization and Computer Graphics}, vol.~25,
  no.~1, pp. 43--53, 2018.

\bibitem{brodlie2004distributed}
K.~W. Brodlie, D.~A. Duce, J.~R. Gallop, J.~P. Walton, and J.~D. Wood,
  ``{Distributed and Collaborative Visualization},'' \emph{Computer Graphics
  Forum}, vol.~23, no.~2, pp. 223--251, 2004.

\bibitem{dasgupta2014opportunities}
A.~Dasgupta, E.~Maguire, A.~Abdul-Rahman, and M.~Chen, ``{Opportunities and
  Challenges for Privacy-Preserving Visualization of Electronic Health Record
  Data},'' in \emph{Proceedings of the IEEE VIS Workshop on Visualization of
  Electronic Health Records}, 2014.

\bibitem{wang2017utility}
X.~Wang, J.-K. Chou, W.~Chen, H.~Guan, W.~Chen, T.~Lao, and K.-L. Ma, ``{A
  Utility-aware Visual Approach for Anonymizing Multi-attribute Tabular
  Data},'' \emph{IEEE Transactions on Visualization and Computer Graphics},
  vol.~24, no.~1, pp. 351--360, 2017.

\bibitem{wang2018graphprotector}
X.~Wang, W.~Chen, J.-K. Chou, C.~Bryan, H.~Guan, W.~Chen, R.~Pan, and K.-L. Ma,
  ``{GraphProtector: A Visual Interface for Employing and Assessing Multiple
  Privacy Preserving Graph Algorithms},'' \emph{IEEE Transactions on
  Visualization and Computer Graphics}, vol.~25, no.~1, pp. 193--203, 2018.

\bibitem{dasgupta2011adaptive}
A.~Dasgupta and R.~Kosara, ``{Adaptive Privacy-Preserving Visualization Using
  Parallel Coordinates},'' \emph{IEEE Transactions on Visualization and
  Computer Graphics}, vol.~17, no.~12, pp. 2241--2248, 2011.

\bibitem{wang2016survey}
X.-M. Wang, T.-Y. Zhang, Y.-X. Ma, J.~Xia, and W.~Chen, ``A survey of visual
  analytic pipelines,'' \emph{Journal of Computer Science and Technology},
  vol.~31, no.~4, pp. 787--804, 2016.

\bibitem{agrawal2002hippocratic}
R.~Agrawal, J.~Kiernan, R.~Srikant, and Y.~Xu, ``{Hippocratic Databases},'' in
  \emph{Proceedings of the 28th International Conference on Very Large
  Databases}, 2002, pp. 143--154.

\bibitem{sweeney2002k}
L.~Sweeney, ``{\textit{k}-anonymity: A model for protecting privacy},''
  \emph{International Journal of Uncertainty, Fuzziness and Knowledge-Based
  Systems}, vol.~10, no.~5, pp. 557--570, 2002.

\bibitem{machanavajjhala2007diversity}
A.~Machanavajjhala, D.~Kifer, J.~Gehrke, and M.~Venkitasubramaniam,
  ``{\textit{l}-Diversity: Privacy Beyond \textit{k}-Anonymity},'' \emph{ACM
  Transactions on Knowledge Discovery from Data}, vol.~1, no.~1, p.~3, 2007.

\bibitem{li2007t}
N.~Li, T.~Li, and S.~Venkatasubramanian, ``{\textit{t}-Closeness: Privacy
  Beyond \textit{k}-Anonymity and \textit{l}-Diversity},'' in \emph{Proceedings
  of the IEEE 23rd International Conference on Data Engineering}, 2007, pp.
  106--115.

\bibitem{cao2012publishing}
J.~Cao and P.~Karras, ``Publishing microdata with a robust privacy guarantee,''
  in \emph{Proceedings of the VLDB Endowment}, 2012, pp. 1388 -- 1399.

\bibitem{soria2013differential}
J.~Soria-Comas and J.~Domingo-Ferrert, ``Differential privacy via t-closeness
  in data publishing,'' in \emph{2013 Eleventh Annual Conference on Privacy,
  Security and Trust}.\hskip 1em plus 0.5em minus 0.4em\relax IEEE, 2013, pp.
  27--35.

\bibitem{chou2016privacy}
J.-K. Chou, Y.~Wang, and K.-L. Ma, ``{Privacy Preserving Event Sequence Data
  Visualization using a Sankey Diagram-like Representation},'' in
  \emph{Proceedings of the SIGGRAPH ASIA Symposium on Visualization}, 2016, pp.
  1--8.

\bibitem{chou2019privacy}
------, ``{Privacy Preserving Visualization: A Study on Event Sequence Data},''
  \emph{Computer Graphics Forum}, vol.~38, no.~1, pp. 340--355, 2019.

\bibitem{dasguptaguess}
A.~Dasgupta, R.~Kosara, and M.~Chen, ``{\textit{Guess Me If You Can}: A Visual
  Uncertainty Model for Transparent Evaluation of Disclosure Risks in
  Privacy-Preserving Data Visualization},'' in \emph{Proceedings of the IEEE
  Symposium on Visualization for Cyber Security}, 2019.

\bibitem{chen2020federated}
W.~Chen, Y.~Wei, Z.~Wang, S.~Zhou, B.~Lin, and Z.~Zhou, ``Federated
  visualization: A privacy-preserving strategy for decentralized
  visualization,'' \emph{arXiv preprint arXiv:2007.15227}, 2020.

\bibitem{lefebvre2006perfect}
S.~Lefebvre and H.~Hoppe, ``{Perfect Spatial Hashing},'' \emph{ACM Transactions
  on Graphics}, vol.~25, no.~3, pp. 579--588, 2006.

\bibitem{xie2014visualizing}
J.~Xie, H.~Yu, and K.-L. Ma, ``{Visualizing Large 3D Geodesic Grid Data with
  Massively Distributed GPUs},'' in \emph{Proceedings of the IEEE 4th Symposium
  on Large Data Analysis and Visualization}, 2014, pp. 3--10.

\bibitem{shih2016parallel}
M.~Shih, S.~Rizzi, J.~Insley, T.~Uram, V.~Vishwanath, M.~Hereld, M.~E. Papka,
  and K.-L. Ma, ``{Parallel Distributed, GPU-Accelerated, Advanced Lighting
  Calculations for Large-Scale Volume Visualization},'' in \emph{Proceedings of
  the IEEE 6th Symposium on Large Data Analysis and Visualization}, 2016, pp.
  47--55.

\bibitem{meidiana2019topology}
A.~Meidiana, S.-H. Hong, J.~Huang, P.~Eades, and K.-L. Ma, ``{Topology-Based
  Spectral Sparsification},'' in \emph{Proceedings of the IEEE 9th Symposium on
  Large Data Analysis and Visualization}, 2019, pp. 73--82.

\bibitem{arleo2017graphray}
A.~Arleo, O.-H. Kwon, and K.-L. Ma, ``{GraphRay: Distributed Pathfinder Network
  Scaling},'' in \emph{Proceedings of the IEEE 7th Symposium on Large Data
  Analysis and Visualization}, 2017, pp. 74--83.

\bibitem{konevcny2016federated}
J.~Kone{\v{c}}n{\`y}, H.~B. McMahan, F.~X. Yu, P.~Richt{\'a}rik, A.~T. Suresh,
  and D.~Bacon, ``{Federated Learning: Strategies for Improving Communication
  Efficiency},'' \emph{arXiv preprint arXiv:1610.05492}, 2016.

\bibitem{li2019federated}
T.~Li, A.~K. Sahu, A.~Talwalkar, and V.~Smith, ``{Federated Learning:
  Challenges, Methods, and Future Directions},'' \emph{arXiv preprint
  arXiv:1908.07873}, 2019.

\bibitem{wei2019multi}
X.~Wei, Q.~Li, Y.~Liu, H.~Yu, T.~Chen, and Q.~Yang, ``Multi-agent visualization
  for explaining federated learning.'' in \emph{Proceedings of the
  Twenty-Eighth IJCAI}, 2019, pp. 6572--6574.

\bibitem{konevcny2016federatedoptimization}
J.~Kone{\v{c}}n{\`y}, H.~B. McMahan, D.~Ramage, and P.~Richt{\'a}rik,
  ``{Federated Optimization: Distributed Machine Learning for On-Device
  Intelligence},'' \emph{arXiv preprint arXiv:1610.02527}, 2016.

\bibitem{mcmahan2017learning}
H.~B. McMahan, D.~Ramage, K.~Talwar, and L.~Zhang, ``{Learning Differentially
  Private Recurrent Language Models},'' \emph{arXiv preprint arXiv:1710.06963},
  2017.

\bibitem{geyer2017differentially}
R.~C. Geyer, T.~Klein, and M.~Nabi, ``{Differentially Private Federated
  Learning: A Client Level Perspective},'' \emph{arXiv preprint
  arXiv:1712.07557}, 2017.

\bibitem{bonawitz2016practical}
K.~Bonawitz, V.~Ivanov, B.~Kreuter, A.~Marcedone, H.~B. McMahan, S.~Patel,
  D.~Ramage, A.~Segal, and K.~Seth, ``{Practical Secure Aggregation for
  Federated Learning on User-Held Data},'' \emph{arXiv preprint
  arXiv:1611.04482}, 2016.

\bibitem{huang2019patient}
L.~Huang, A.~L. Shea, H.~Qian, A.~Masurkar, H.~Deng, and D.~Liu, ``{Patient
  clustering improves efficiency of federated machine learning to predict
  mortality and hospital stay time using distributed electronic medical
  records},'' \emph{Journal of Biomedical Informatics}, vol.~99, p. 103291,
  2019.

\bibitem{huang2018loadaboost}
L.~Huang, Y.~Yin, Z.~Fu, S.~Zhang, H.~Deng, and D.~Liu, ``{LoADaBoost:
  Loss-Based AdaBoost Federated Machine Learning on medical data},''
  \emph{arXiv preprint arXiv:1811.12629}, 2018.

\bibitem{liu2020fedvision}
Y.~Liu, A.~Huang, Y.~Luo, H.~Huang, Y.~Liu, Y.~Chen, L.~Feng, T.~Chen, H.~Yu,
  and Q.~Yang, ``{FedVision: An Online Visual Object Detection Platform Powered
  by Federated Learning},'' \emph{arXiv preprint arXiv:2001.06202}, 2020.

\bibitem{yang2019federated}
Q.~Yang, Y.~Liu, T.~Chen, and Y.~Tong, ``{Federated Machine Learning: Concept
  and Applications},'' \emph{ACM Transactions on Intelligent Systems and
  Technology}, vol.~10, no.~2, pp. 1--19, 2019.

\bibitem{ramsundar2018tensorflow}
B.~Ramsundar and R.~B. Zadeh, \emph{{TensorFlow for Deep Learning: From Linear
  Regression to Reinforcement Learning}}, 2018.

\bibitem{dwork2006calibrating}
C.~Dwork, F.~McSherry, K.~Nissim, and A.~Smith, ``Calibrating noise to
  sensitivity in private data analysis,'' in \emph{Theory of cryptography
  conference}.\hskip 1em plus 0.5em minus 0.4em\relax Springer, 2006, pp.
  265--284.

\bibitem{mei2019rsatree}
H.~Mei, W.~Chen, Y.~Wei, Y.~Hu, S.~Zhou, B.~Lin, Y.~Zhao, and J.~Xia,
  ``{RSATree: Distribution-Aware Data Representation of Large-Scale Tabular
  Datasets for Flexible Visual Query},'' \emph{IEEE Transactions on
  Visualization and Computer Graphics}, vol.~26, no.~1, pp. 1161--1171, 2019.

\bibitem{zhao2018federated}
Y.~Zhao, M.~Li, L.~Lai, N.~Suda, D.~Civin, and V.~Chandra, ``{Federated
  Learning with Non-IID Data},'' \emph{arXiv preprint arXiv:1806.00582}, 2018.

\bibitem{wood2010visualisation}
J.~Wood, J.~Dykes, and A.~Slingsby, ``Visualisation of origins, destinations
  and flows with od maps,'' \emph{The Cartographic Journal}, vol.~47, no.~2,
  pp. 117--129, 2010.

\end{thebibliography}



%


\begin{IEEEbiography}[{\includegraphics[width=1in,height=1.25in,clip,keepaspectratio]{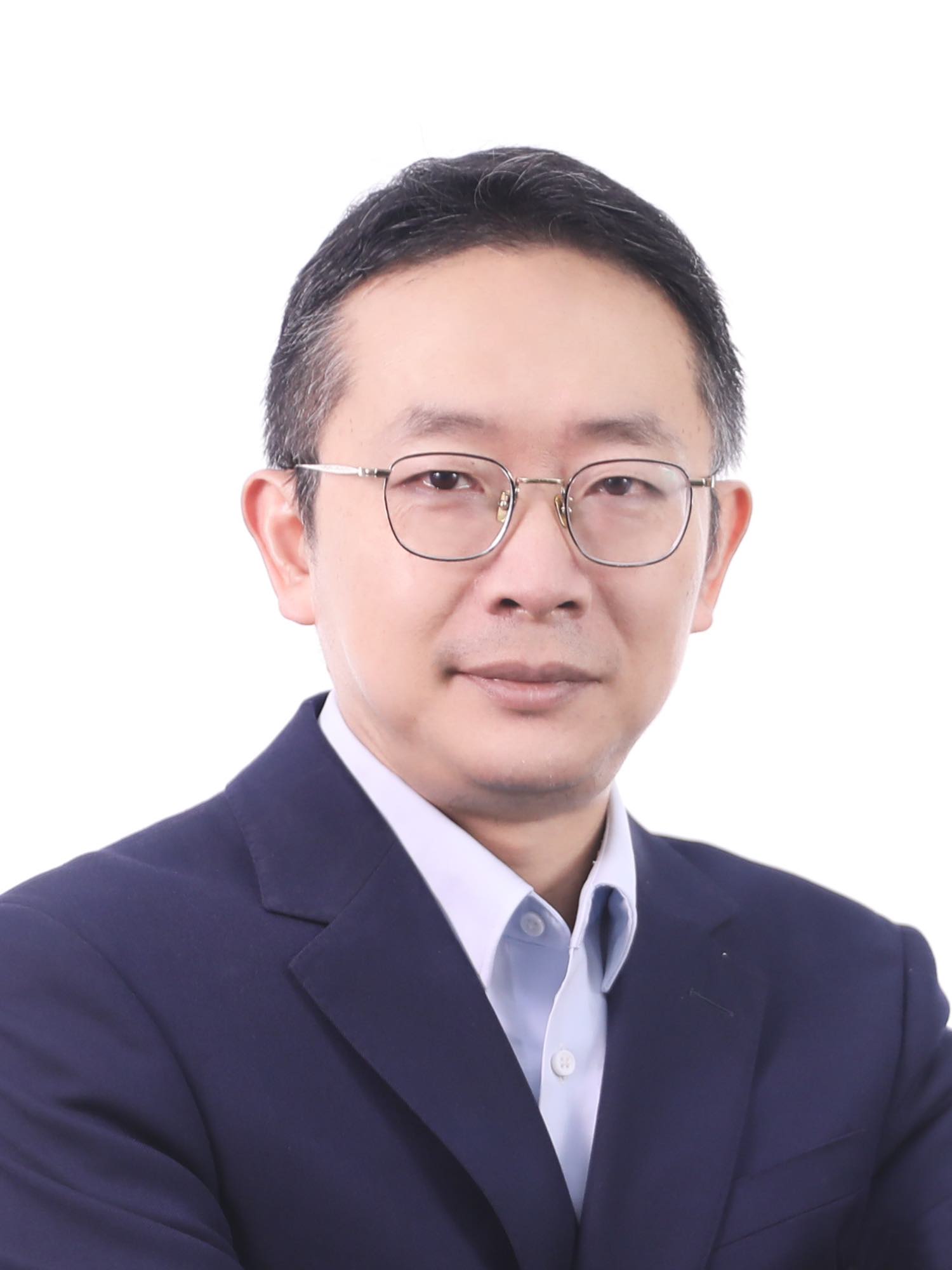}}]{Wei Chen} is a professor in the State Key Lab of CAD\&CG, Zhejiang University. His research interests include visualization and visual analysis, and has published more than 70 IEEE/ACM Transactions and IEEE VIS papers. He actively served as guest or associate editors of the ACM Transactions on Intelligent System and Technology, the IEEE Computer Graphics and Applications and Journal of Visualization. 
\end{IEEEbiography}

\begin{IEEEbiography}[{\includegraphics[width=1in,height=1.25in,clip,keepaspectratio]{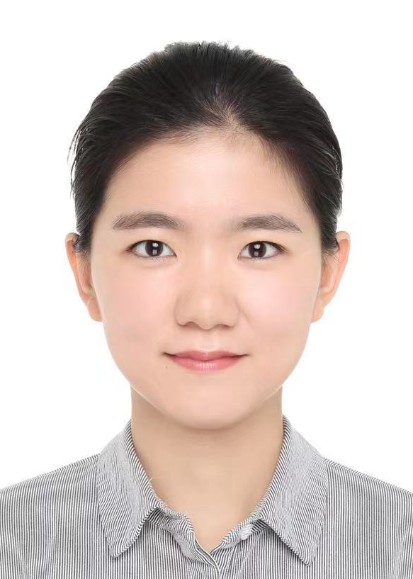}}]{Yating Wei} is a Ph.D. student in the State Key Lab of CAD\&CG at Zhejiang University, Hangzhou. She earned the B.S. degree in software engineering from Central South University in 2017. Her research interests are visual analytics and perceptual consistency. 
\end{IEEEbiography}

\begin{IEEEbiography}[{\includegraphics[width=1in,height=1.25in,clip,keepaspectratio]{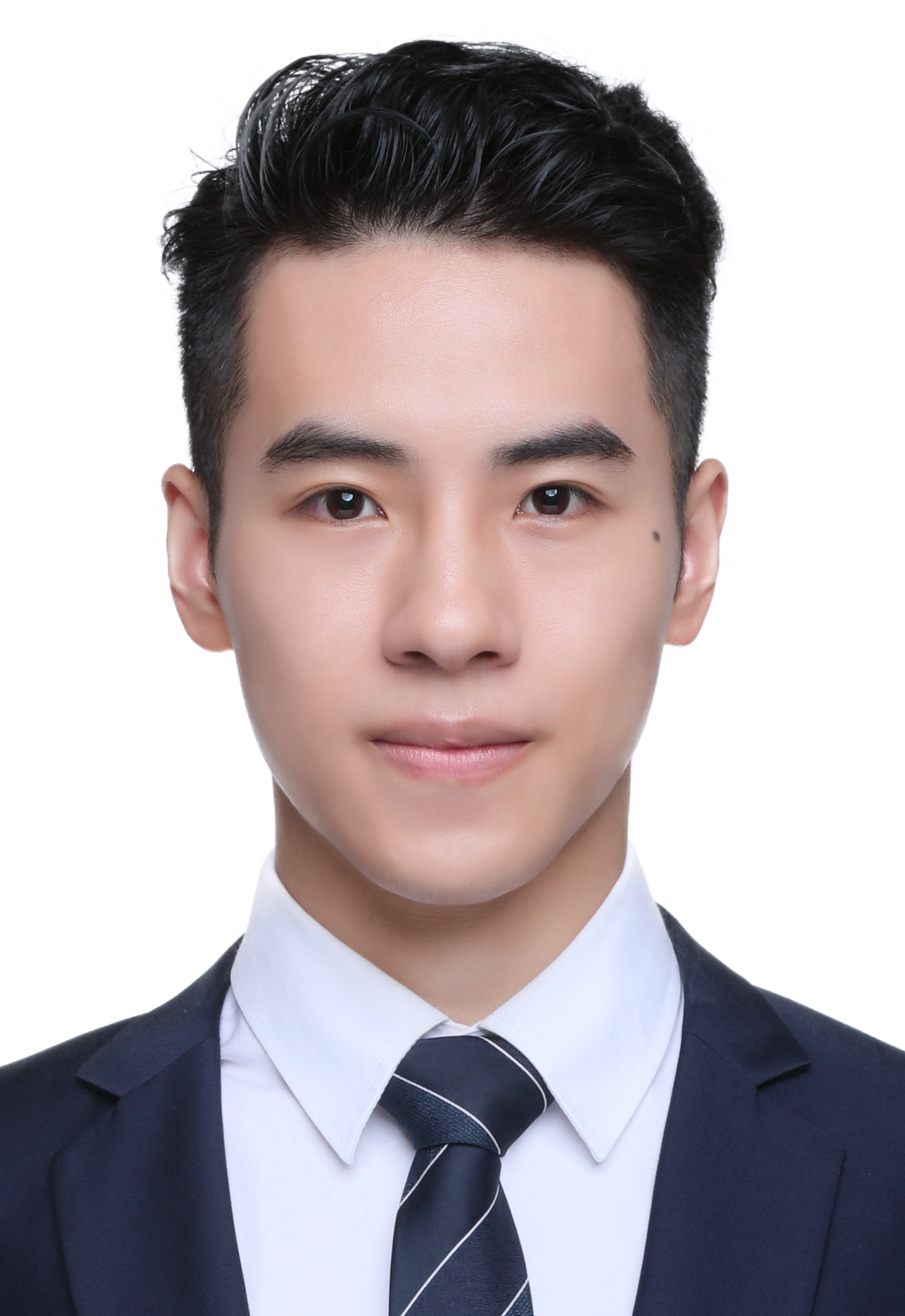}}]{Zhiyong Wang} is a postgraduate in the State Key Lab of CAD\&CG at Zhejiang University, Hangzhou. His research interests are visual analytics. 
\end{IEEEbiography}

\begin{IEEEbiography}[{\includegraphics[width=1in,height=1.25in,clip,keepaspectratio]{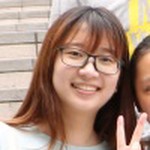}}]
{Shuyue Zhou} received the master degree at 2021 at Zhejiang University, China. She is now working in Alibaba Group, Hangzhou, China. Her research interests include information visualization and visual analytics. 
\end{IEEEbiography}

\begin{IEEEbiography}[{\includegraphics[width=1in,height=1.25in,clip,keepaspectratio]{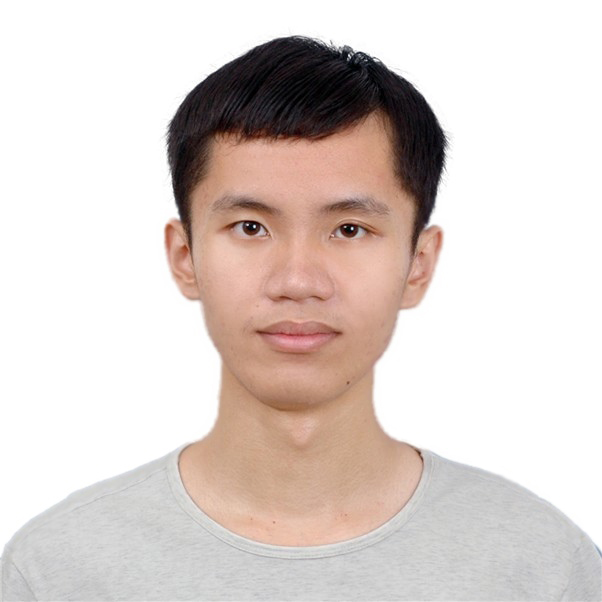}}]
{Bingru Lin} received the master degree at 2020 at Zhejiang University, China. He is now working in Alibaba Group, Hangzhou, China. His research interests include information visualization and visual analytics. 
\end{IEEEbiography}

\begin{IEEEbiography}[{\includegraphics[width=1in,height=1.25in,clip,keepaspectratio]{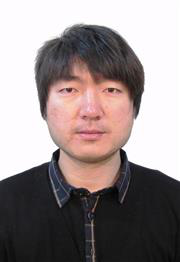}}]
{Zhiguang Zhou} is a professor in the School of Information at the Zhejiang University of Finance and Economics. His research interests include information visualization, visual analytics. Zhou has a doctor degree in computer science from Zhejiang University. 
\end{IEEEbiography}






\end{document}